\documentclass[prd,showpacs,showkeys,nofootinbib,preprint]{revtex4-1}
\usepackage{amsmath,amsfonts,amssymb,color}
\usepackage{bm}
\usepackage{pgfplots}
\usepackage[colorlinks=true,urlcolor=blue,linkcolor=blue,citecolor=blue]{hyperref}

\usepackage{amsmath,amssymb,color,graphics,graphicx}
\usepackage{bm}
\usepackage{hyperref} 

\definecolor{dark-green}{rgb}{0,0.7,0}
\definecolor{dark-blue}{rgb}{0,0.2,0.5}
\definecolor{med-blue}{rgb}{0,0.7,1}
\definecolor{mblue}{rgb}{0,0.2,1}
\definecolor{cnc}{rgb}{0.8,0,0}
\definecolor{light-red}{rgb}{1,0.8,0.8}
\definecolor{dark-yellow}{rgb}{1,0.8,0}
\definecolor{light-blue}{rgb}{0.8,0.9,1}
\definecolor{verylight-blue}{rgb}{0.93,0.95,1}
\definecolor{light-yellow}{rgb}{1,0.9,0.8}
\definecolor{grey}{gray}{0.88}

\def\a{\alpha}
\def\b{\beta}

\def\d{\delta}
\def\g{\gamma}

\usepackage{amsfonts}

\newcommand{\aR}{\nearrow\!\!\!\!\!\!\!R}
\newcommand{\aOR}{\nearrow\!\!\!\!\!\!\!\overline{R}}

\begin{document}

\title{Poincar\'e gauge gravity primer}

\author{Yuri N. Obukhov}

\affiliation{Russian Academy of Sciences, Nuclear Safety Institute, 
B.Tulskaya 52, 115191 Moscow, Russia
\email{obukhov@ibrae.ac.ru}}

\begin{abstract}
We give an introductory overview of the classical Poincar\'e gauge theory of gravity formulated on the spacetime manifold that carries the Riemann-Cartan geometry with nontrivial curvature and torsion. After discussing the basic mathematical structures at an elementary level in the framework of the standard tensor analysis, we formulate the general dynamical scheme of Poincar\'e gauge gravity for the class of Yang-Mills type models, and consider a selected number of physically interesting consequences of this theory.
\end{abstract}

\maketitle

\section{Introduction}

The gauge-theoretic approach in classical field theory has a long history (going back to the early works of Weyl, Cartan, Fock, for an overview see \cite{ORaifeartaigh:2000,Straumann:2020,Hehl:2020}) and it underlies the modern understanding of the nature of the physical interactions \cite{ORaifeartaigh:1978,Mack:1979,Chaichian:1984}. The original Yang-Mills \cite{Yang:1954} treatment of the internal symmetry groups was subsequently extended to the spacetime symmetries by Utiyama \cite{Utiyama:1956}, Sciama \cite{Sciama:1962} and Kibble \cite{Kibble:1961}. The detailed review of the development of the gauge approach in gravity theory and the corresponding mathematical structures can be found in \cite{Hehl:1976kj,Erice:1979,Hehl:EinsteinVolume,Trautman:1982,Nester:1984,PRs,Shapiro,Blagojevic:2002,Blagojevic:2013,Nester:2016,Mielke,Ponomarev}. It is worthwhile to mention that the book \cite{Ponomarev} provides an essentially complete bibliography on this subject. Here we do not intend to present an exhaustive and comprehensive review of the Poincar\'e gauge gravity theory, and give a rather concise and elementary introduction into the subject. One may view this paper as a continuation of the earlier work \cite{Hehl:deBroglie,Obukhov:2006,Obukhov:2018}. 

The Poincar\'e gauge (PG) gravity is a natural extension of Einstein's general relativity (GR) theory. Being based on gauge-theoretic principles, it takes into account the spin (commonly viewed as a microstructural property of matter) as an additional physical source of the gravitational field on an equal footing with the energy and momentum (naturally viewed as macroscopic properties of matter). The corresponding spacetime structure is then adequately described by the Riemann-Cartan geometry with curvature and torsion. From the mathematical point of view, the PG formalism arises as a special case of the metric-affine gravity (MAG) theory \cite{PRs} that provides a unified framework for the study of alternative theories based on post-Riemannian geometries \cite{Schrodinger,Schouten:1954}. Other special cases of MAG include the geometries of Riemann of GR \cite{Einstein:meaning}, Weyl \cite{Weyl:1923}, Weitzenb\"ock \cite{Weitzenboeck:1923}, etc.

The Poincar\'e gauge gravity occupies a prominent place in the colorful landscape of modified gravitational theories that generalize or extend the physical and mathematical structure of Einstein's GR. Among such theories it is worthwhile to highlight the large classes of $f(R)$ and $f(T)$ models, and of theories with nonminimal coupling to matter, developed mainly in the context of relativistic cosmology, see \cite{Harko:2014:1,Schmidt:2007,Bertolami:etal:2007,Straumann:2018,Nojiri:2011,Faraoni:2011,Bahamonde:2021}. The so-called Palatini approach represents another class of widely discussed theories in which metric and connection are treated as independent variables in the action principle \cite{Hehl:1978,Hehl:1981,Sotiriou:2010}. 

Our basic notation and conventions are consistent with \cite{PRs,Birkbook}. In particular, Greek indices $\alpha, \beta, \dots = 0, \dots, 3$, denote the anholonomic components (for example, of an orthonormal frame $e_\alpha$), while the Latin indices $i,j,\dots =0,\dots, 3$, label the holonomic components (e.g., the world coordinate basis $e_i$). Spatial components are numbered by Latin indices from the beginning of the alphabet $a, b, \dots = 1, 2, 3$. To distinguish separate holonomic components from the anholonomic ones, we put hats over the latter indices: e.g., $e_\alpha = \{e_{\hat{0}}, e_{\hat{1}}, e_{\hat{2}}, e_{\hat{3}}\}$ vs. $e_i = \{e_0, e_1, e_2, e_3\}$. The totally antisymmetric Levi-Civita is denoted $\eta_{ijkl}$. The Minkowski metric is $g_{\alpha\beta} = {\rm diag}(c^2,-1,-1,-1)$. All the objects related to the parity-odd sector (coupling constants, irreducible pieces of the curvature, etc.) are marked by an overline, to distinguish them from the corresponding parity-even objects.

\section{Riemann-Cartan geometry}

We model spacetime as a four-dimensional smooth manifold $M$, and leaving aside the global (topological) aspects, we focus only on local issues. The local coordinates $x^i$, $i = 0,1,2,3$, are introduced in the neighborhood of an arbitrary point of the spacetime manifold. The geometrical (gravitational) and physical (material) variables are then fields of different nature (both tensors and nontensors) over the spacetime. They are characterized by their components and transformation properties under local diffeomorphisms $x^i\rightarrow x'^i(x^k)$. An infinitesimal diffeomorphism $x^i\rightarrow x^i + \delta x^i$, 
\begin{equation}
\delta x^i = \xi^i(x),\label{dex}
\end{equation}
is thus parametrized by the four arbitrary functions $\xi^i(x)$.

\subsection{Geometrical structures}

In the framework of what can be quite generally called an Einsteinian approach (with the principles of equivalence and general coordinate covariance as the cornerstones), the gravitational phenomena are described by the two fundamental geometrical structures on a spacetime manifold: the metric $g_{ij}$ and connection $\Gamma_{ki}{}^j$. As Einstein himself formulated \cite{Einstein:meaning}, the crucial achievement of his theory was the elimination of the notion of inertial systems as preferred ones among all possible coordinate systems. 

From the geometrical point of view, the metric introduces lengths and angles of vectors, and thereby determines the distances (intervals) between points on the spacetime manifold. The connection introduces the notion of parallel transport and defines the covariant differentiation $\nabla_k$ of tensor fields. In the metric-affine theory of gravity, the connection is not necessarily symmetric and compatible with the metric. Under infinitesimal diffeomorphisms (\ref{dex}), these geometrical variables transform as 
\begin{align}
\delta g_{ij} &= -\,(\partial_i\xi^k)\,g_{kj} - (\partial_j\xi^k)\,g_{ik},\label{dgij}\\
\delta\Gamma_{ki}{}^j &= -\,(\partial_k\xi^l)\,\Gamma_{li}{}^j - (\partial_i\xi^l)\,\Gamma_{kl}{}^j + (\partial_l\xi^j)\,\Gamma_{ki}{}^l - \partial^2_{ki}\xi^j.\label{dG}
\end{align}

The {\it Riemann-Cartan geometry} of a spacetime manifold is characterized by two tensors: the curvature and the torsion which are defined \cite{Schouten:1954} as
\begin{align}
R_{kli}{}^j &:= \partial_k\Gamma_{li}{}^j - \partial_l\Gamma_{ki}{}^j + \Gamma_{kn}{}^j \Gamma_{li}{}^n - \Gamma_{ln}{}^j\Gamma_{ki}{}^n,\label{curv}\\
T_{kl}{}^i &:= \Gamma_{kl}{}^i - \Gamma_{lk}{}^i,\label{tors}
\end{align}
whereas the nonmetricity vanishes:
\begin{equation}\label{nonmet}
Q_{kij} := -\,\nabla_kg_{ij} = - \partial_kg_{ij} + \Gamma_{ki}{}^lg_{lj} + \Gamma_{kj}{}^lg_{il} = 0.
\end{equation}
The curvature and the torsion tensors determine the commutator of the covariant derivatives. For a tensor $A^{i_1 \dots i_p}{}_{j_1 \dots j_q}$ of arbitrary rank and index structure: 
\begin{eqnarray}
&& (\nabla_k\nabla_l - \nabla_l\nabla_k) A^{i_1 \dots i_p}{}_{j_1 \dots j_q} = -\,T_{kl}{}^n\nabla_n A^{i_1 \dots i_p}{}_{j_1 \dots j_q} \nonumber \\
&& + \sum^{p}_{r=1} R_{kln}{}^{i_r} A^{i_1 \dots n \dots i_p}{}_{j_1 \dots j_q} - \sum^{q}_{r=1}R_{klj_r}{}^{n} A^{i_1 \dots i_p}{}_{j_1 \dots n \dots j_q}. \label{commutator}
\end{eqnarray}
By applying the covariant derivative $\nabla_l$ to the metricity condition (\ref{nonmet}), and evaluating the commutator of covariant derivatives, we find
\begin{equation}
R_{lk(ij)} = 0,\label{Dnonmet}
\end{equation}
i.e., the curvature tensor is skew-symmetric in both pairs of its indices. 

The Riemannian connection $\widetilde{\Gamma}_{kj}{}^i$ is uniquely determined by the conditions of vanishing torsion and nonmetricity which yield explicitly 
\begin{equation}\label{Chr}
\widetilde{\Gamma}_{kj}{}^i = {\frac 12}g^{il}(\partial_jg_{kl} + \partial_kg_{lj} - \partial_lg_{kj}).
\end{equation}
Here and in the following, a tilde over a symbol denotes a Riemannian object (such as the curvature tensor) or a Riemannian operator (such as the covariant derivative) constructed from the Christoffel symbols (\ref{Chr}). The deviation of the Riemann-Cartan geometry from the Riemannian one is then conveniently described by the {\it contortion} tensor 
\begin{equation}
K_{kj}{}^i := \widetilde{\Gamma}_{kj}{}^i - \Gamma_{kj}{}^i.\label{dist}
\end{equation}
The system (\ref{tors}) and (\ref{nonmet}) allows to find the contortion tensor in terms of the torsion:
\begin{equation}\label{NTQ}
K_{kj}{}^i = -\,{\frac 12}(T_{kj}{}^i + T^i{}_{kj} + T^i{}_{jk})\,.
\end{equation}
From this we can check the skew symmetry in the two last indices, $K_{k(ij)} = 0$, which is also seen directly when we use (\ref{dist}) in (\ref{nonmet}). Furthermore, combining (\ref{dist}) with (\ref{tors}) one can express the torsion tensor in terms of the contortion,
\begin{equation}
T_{kj}{}^i = -\,2K_{[kj]}{}^i.\label{TN}
\end{equation}
Substituting (\ref{dist}) into (\ref{curv}), we find the relation between the non-Riemannian and the Riemannian curvature tensors
\begin{equation}
R_{kli}{}^j = \widetilde{R}_{kli}{}^j - \widetilde{\nabla}_kK_{li}{}^j
+ \widetilde{\nabla}_lK_{ki}{}^j + K_{kn}{}^jK_{li}{}^n - K_{ln}{}^jK_{ki}{}^n.\label{RRN}
\end{equation}

Applying the covariant derivative to (\ref{curv})-(\ref{nonmet}) and antisymmetrizing, we derive the Bianchi identities \cite{Schouten:1954}:
\begin{align}
\nabla_{[n}R_{kl]i}{}^j &= T_{[kl}{}^m R_{n]mi}{}^j\,,\label{Dcurv}\\
\nabla_{[n}T_{kl]}{}^i &= R_{[kln]}{}^i + T_{[kl}{}^m T_{n]m}{}^i\,.\label{Dtors}
\end{align}

\subsection{Special cases}

When the torsion vanishes,
\begin{equation}
T_{ij}{}^k = 0,\label{Rie}
\end{equation}
the Riemann-Cartan spacetime reduces to the Riemannian geometry of Einstein's GR, which is characterized by the curvature $\widetilde{R}_{kli}{}^j$ constructed from the Christoffel symbols (\ref{Chr}). In this case, the connection is no longer an independent dynamical variable.

Quite remarkably, the vanishing curvature condition
\begin{equation}
R_{kli}{}^j = 0,\label{Wei}
\end{equation}
also produces a meaningful spacetime structure. This is known as the {\it Weitzenb\"ock geometry} \cite{Weitzenboeck:1923} which is characterized by the property of a distant parallelism \cite{Weitzenboeck:1928,Einstein:1928}: the result of a parallel transport of a vector from a point $x$ to a point $y$ does not depend on a path along which it is transported. The  Weitzenb\"ock geometry underlies another interesting gauge gravity theory which is based on the group of spacetime translations \cite{Hay1,Cho:1975dh,Nitsch:1979qn,Hay2,Obukhov:2002tm,Pereira:2019,Koivisto:2019,Aldrovandi:2013,Maluf:2012,Maluf:2013,Itin:2001,Hehl:2016glb,Itin:2018dru,Itin:2016nxk,illumi}. 

\begin{figure}
\centering
\includegraphics[width=0.75\textwidth]{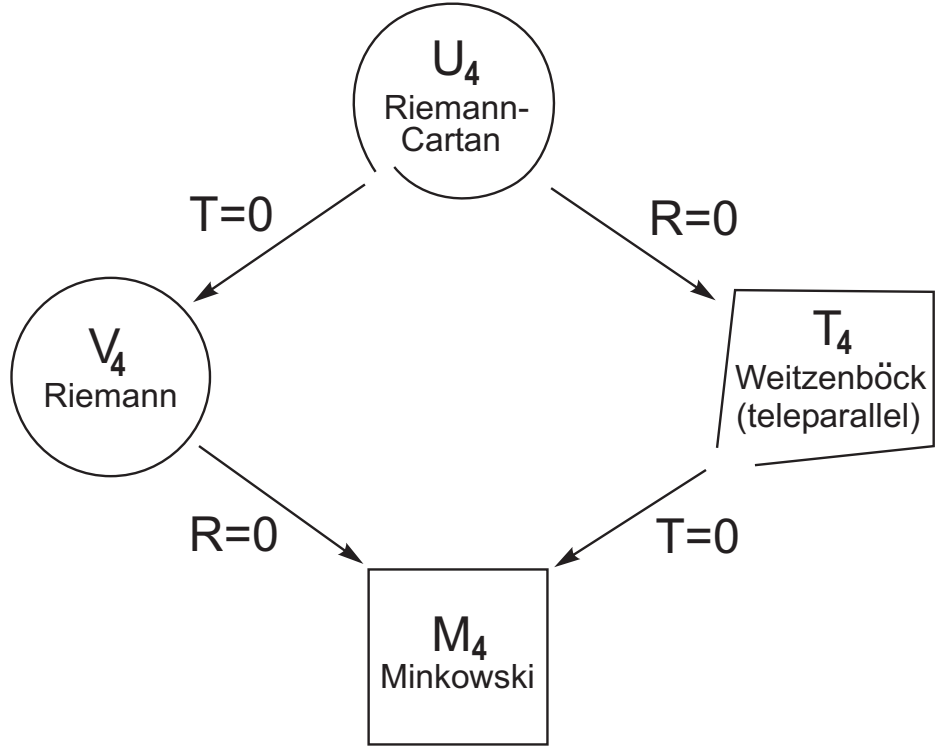}
\caption{A Riemann-Cartan space $U_4$ with torsion $T$ and curvature $R$ and its different
limits (nonmetricity vanishes: $Q_{kij}:=-\nabla_kg_{ij}=0$), see \cite{Blagojevic:2013}, p.174.}\label{RCgeom}
\end{figure}

When both conditions (\ref{Rie}) and (\ref{Wei}) are satisfied, the spacetime reduces to a flat Minkowski geometry. Fig.~\ref{RCgeom} summarizes the landscape of special cases of the Riemann-Cartan geometry. 

\subsection{Local Lorentz structures (frame formalism)}\label{tetrad}

The components of geometrical objects above are defined with respect to a coordinate basis $e_i$ of the tangent space, which is composed of four vectors tangential to the coordinate lines $x^i$. Under the change of the local coordinates $x^i\rightarrow x'^i(x^k)$ the coordinate basis transforms as
\begin{equation}
e'_k = {\frac {\partial x^i}{\partial x'^k}}\,e_i,\label{Ctra}
\end{equation}
and this yields the corresponding transformation of the metric, connection, and other geometrical objects. The vectors of the coordinate basis $e_i$ are neither unit, nor orthogonal; their lengths and mutual angles are encoded in the components of the metric via the definition of the scalar product: $g_{ij} = (e_i, e_j)$.

It is then reasonable to introduce, at every point of the spacetime manifold, a local orthonormal frame $e_\alpha$ for which the scalar product $(e_\alpha, e_\beta) = g_{\alpha\beta}$ is equal to the Minkowski metric. To distinguish the local Lorentz frame from the coordinate basis, we label their legs by Greek letters instead of the Latin ones. Decomposing $e_\alpha$ with the respect to $e_i$, 
\begin{equation}
e_\alpha = e_\alpha^i\,e_i\,,
\end{equation}
we find the frame (tetrad, or vierbein) components $e_\alpha^i(x)$. Repeating the same for the coframe (basis of the cotangent space), we eventually obtain the inverse orthonormal coframe with components $e^\alpha_i(x)$. By construction we have
\begin{equation}
g_{ij} = e^\alpha_i e^\beta_jg_{\alpha\beta},\label{geeg}
\end{equation}
which explains why the coframe is sometimes called a ``square root'' of the metric. Normally, a local Lorentz frame is anholonomic since it cannot be constructed from partial derivatives $e^\alpha_i\neq \partial_if^\alpha$ of some functions $f^\alpha(x)$, and this is measured by the anholonomity object
\begin{equation}
C_{ij}{}^\alpha = \partial_ie^\alpha_j - \partial_je^\alpha_i.\label{anhol}
\end{equation}

In order to describe the parallel transport of vectors with respect to local Lorentz frames, one needs to transform the connection components accordingly:
\begin{equation}\label{GG1}
\Gamma_{k\alpha}{}^\beta = e_\alpha^ie^\beta_j\Gamma_{ki}{}^j + e^\beta_i\partial_ke_\alpha^i.
\end{equation}
It is worthwhile to notice that we can rewrite this as
\begin{equation}\label{GG0}
\partial_ke^\alpha_i + \Gamma_{k\beta}{}^\alpha e^\beta_i - \Gamma_{ki}{}^ie^\alpha_j = 0,
\end{equation}
which in the literature is sometimes called a ``postulate'' of the vanishing of the ``total'' covariant derivative of the coframe. This is an unfortunate misunderstanding (sadly, a widely spread one). Both relations are not postulated, they merely describe a transformation law of the connection. The inverse transformation is straightforwardly derived from (\ref{GG1}):
\begin{equation}\label{GG2}
\Gamma_{ki}{}^j = e_\beta^je^\alpha_i\Gamma_{k\alpha}{}^\beta + e^j_\beta\partial_ke^\beta_i.
\end{equation}

By definition (\ref{Ctra}), the tetrad relates the world coordinate and the local Lorentz components of geometrical objects, e.g., $V^\alpha = e^\alpha_iV^i$. By making use of (\ref{GG1})-(\ref{GG2}), we then also relate the covariant derivatives: $e^\alpha_j\nabla_iV^j = D_iV^\alpha = \partial_iV^\alpha + \Gamma_{i\beta}{}^\alpha V^\beta$. It is convenient to distinguish notationally the covariant derivative in the world coordinates $\nabla_i$ from the covariant derivative $D_i$ in the local Lorentz frames.

In particular, recasting the metricity condition (\ref{nonmet}) into the local Lorentz disguise we can demonstrate the skew symmetry of the local Lorentz connection:
\begin{equation}\label{nonmet1}
-\,e_\alpha^ie_\beta^j\nabla_kg_{ij} = -\,D_kg_{\alpha\beta} = \Gamma_{k\alpha}{}^\gamma g_{\gamma\beta}
+ \Gamma_{k\beta}{}^\gamma g_{\alpha\gamma} = \Gamma_{k\alpha\beta} + \Gamma_{k\beta\alpha} = 0.
\end{equation}

\subsection{Symmetries in Riemann-Cartan space: generalized Killing vectors}\label{killing_sec}

As is well known, symmetries of a Riemannian spacetime are generated by Killing vector fields. Each such field defines a so-called {\it motion} of the spacetime manifold, that is a diffeomorphism which preserves the metric $g_{ij}$. The Lie derivative ${\cal L}_\zeta$ is defined along any vector field $\zeta^i$ and it maps tensors into tensors of the same rank. Let us recall the explicit form of the Lie derivative of the metric and connection \cite{Lichnerowicz:1977,Yano:1955}, which can be derived directly from the transformation laws (\ref{dgij}) and (\ref{dG}):
\begin{align}
{\cal L}_\zeta g_{ij} &= \zeta^k\partial_kg_{ij} + (\partial_i\zeta^k)g_{kj} + (\partial_j\zeta^k)g_{ik},\label{Lm}\\
{\cal L}_\zeta \Gamma_{kj}{}^i &= \zeta^n\partial_n\Gamma_{kj}{}^i + (\partial_k\zeta^n)\Gamma_{nj}{}^i
+ (\partial_j\zeta^n)\Gamma_{kn}{}^i - (\partial_n\zeta^i)\Gamma_{kj}{}^n + \partial_k(\partial_j\zeta^i).\label{Ld}
\end{align}
The latter quantity measures the noncommutativity of the Lie derivative with the covariant derivative
\begin{align}
&({\cal L}_\zeta \nabla_k - \nabla_k{\cal L}_\zeta) A^{i_1 \dots i_p}{}_{j_1 \dots j_q} \nonumber\\
& = \,\sum^{p}_{r=1} ({\cal L}_\zeta\Gamma_{kl}{}^{i_r}) A^{i_1 \dots l \dots i_p}{}_{j_1 \dots j_q}
- \sum^{q}_{r=1} ({\cal L}_\zeta\Gamma_{kj_r}{}^{l}) A^{i_1 \dots j_p}{}_{j_1 \dots l \dots j_q}. \label{commLD}
\end{align}

A vector field $\zeta^i$ is called a {\it Killing vector field} if ${\cal L}_\zeta g_{ij} = 0$. This condition can be recast, using (\ref{Lm}), into an equivalent form
\begin{equation}
\widetilde{\nabla}_i\zeta_j + \widetilde{\nabla}_j\zeta_i = 0,\label{K1}
\end{equation}
which is called a Killing equation. By covariant differentiation with respect to the Riemannian connection, after some algebra we derive from this
\begin{equation}
{\cal L}_\zeta \widetilde{\Gamma}_{ki}{}^j = 0,\qquad {\cal L}_\zeta \widetilde{R}_{kli}{}^j = 0.\label{LR}
\end{equation}
That is, the Lie derivatives along the Killing vector field $\zeta$ vanish for all Riemannian geometrical objects. Moreover, one can show that the same is true for all higher covariant derivatives of the Riemannian curvature tensor \cite{Yano:1955}
\begin{equation}\label{LDR}
{\cal L}_\zeta \bigl(\widetilde{\nabla}_{n_1}\dots \widetilde{\nabla}_{n_N}\widetilde{R}_{kli}{}^j\bigr) = 0.
\end{equation}

Let us generalize the notion of a symmetry to the Riemann-Cartan spacetime. We begin by noticing that for an arbitrary $\lambda_\alpha{}^\beta$ (Greek indices mean that this object is defined with respect to a local Lorentz frame $e^\alpha_i$), we can recast (\ref{Lm}) and (\ref{Ld}) into 
\begin{align} 
{\cal L}_\zeta g_{ij} =& \left({\cal L}_\zeta e^\alpha_i - \lambda_\gamma{}^\alpha e^\gamma_i\right)e^\beta_jg_{\alpha\beta}
+ \bigl({\cal L}_\zeta e^\beta_j - \lambda_\gamma{}^\beta e^\gamma_j\bigr)e^\alpha_ig_{\alpha\beta} \nonumber\\
& + e^\alpha_ie^\beta_j\left({\cal L}_\zeta g_{\alpha\beta} + \lambda_{\alpha\beta} + \lambda_{\beta\alpha}\right),\label{Lm1}\\
{\cal L}_\zeta \Gamma_{ki}{}^j =& \bigl({\cal L}_\zeta e_\alpha^j + \lambda_\alpha{}^\gamma e_\gamma^j\bigr)
D_k e^\alpha_i + e_\alpha^j D_k\bigl({\cal L}_\zeta e^\alpha_i - \lambda_\gamma{}^\alpha e^\gamma_i\bigr)\nonumber\\
& + e_\alpha^j e^\beta_i\left({\cal L}_\zeta \Gamma_{k\beta}{}^\alpha + D_k \lambda_\beta{}^\alpha\right).\label{Ld1}
\end{align}
This is straightforwardly derived from (\ref{geeg}) and (\ref{GG2}) by making use of the standard definitions of the Lie derivatives of the coframe and the local Lorentz connection (which are, geometrically, both covectors), ${\cal L}_\zeta e_k^\alpha = \zeta^i\partial_ie_k^\alpha + (\partial_k\zeta^i)e_i^\alpha$ and ${\cal L}_\zeta \Gamma_{k\beta}{}^\alpha = \zeta^i\partial_i\Gamma_{k\beta}{}^\alpha + (\partial_k\zeta^i)\Gamma_{i\beta}{}^\alpha$, whereas for the world scalar ${\cal L}_\zeta g_{\alpha\beta} = \zeta^i \partial_ig_{\alpha\beta}$.

Accordingly, a natural definition of the symmetry of the Riemann-Cartan manifold can be formulated as the set of conditions
\begin{align}
{\cal L}_\zeta g_{\alpha\beta} &= - \lambda_{\alpha\beta} - \lambda_{\beta\alpha},\label{LRC1}\\
{\cal L}_\zeta e^\alpha_i &= \lambda_\beta{}^\alpha e^\beta_i,\label{LRC2}\\
{\cal L}_\zeta \Gamma_{k\beta}{}^\alpha &= - D_k \lambda_\beta{}^\alpha,\label{LRC3}
\end{align}
where the possible form of $\lambda_\beta{}^\alpha(x)$ is eventually determined by the vector field $\zeta^i(x)$, which is naturally called a {\it generalized Killing vector} of the Riemann-Cartan spacetime. Since the Minkowski metric $g_{\alpha\beta}$ has constant components, (\ref{LRC1}) yields the skew symmetry $\lambda_{\alpha\beta} = -\,\lambda_{\beta\alpha}$. As a result, from (\ref{Lm1}) and (\ref{Ld1}) we find
\begin{align}
{\cal L}_\zeta g_{ij} &= 0,\label{LgMAG}\\
{\cal L}_\zeta \Gamma_{ki}{}^j &= 0.\label{LGMAG}
\end{align}
Thereby, the generalized Killing vector $\zeta^i$ generates a diffeomorphism of the spacetime manifold that is simultaneously an isometry (\ref{LgMAG}) and an isoparallelism (\ref{LGMAG}). 

Combining (\ref{LGMAG}), (\ref{LR}) and (\ref{dist}), we derive ${\cal L}_\zeta K_{ki}{}^j = 0$ for the contortion, and we accordingly conclude that the generalized Killing vector leaves the torsion and the Riemann-Cartan curvature tensors invariant
\begin{equation}
{\cal L}_\zeta T_{ij}{}^k = 0,\qquad {\cal L}_\zeta R_{klj}{}^i = 0.\label{LRMAG}
\end{equation}
It is also straightforward to verify that 
\begin{equation}
{\cal L}_\zeta \bigl(\nabla_{n_1}\dots \nabla_{n_N}T_{ij}{}^k\bigr) = 0,\qquad
{\cal L}_\zeta \bigl(\nabla_{n_1}\dots \nabla_{n_N}R_{klj}{}^i\bigr) = 0\,,\label{LDRMAG} 
\end{equation}
for any number of covariant derivatives of the torsion and the curvature.

\subsection{Matter variables}

Without specializing the discussion of matter to any particular physical field, we can describe matter by a generalized field $\psi^A$. The range of the indices $A,B,\dots$ is not important in our study. However, we do need to know the behavior of the matter field under spacetime diffeomorphisms (\ref{dex}):
\begin{equation}
\delta\psi^A = -\,(\partial_i\xi^j)\,(\sigma_j{}^i)^A{}_B\,\psi^B.\label{dpsiA}
\end{equation}
Here $(\sigma_j{}^i)^A{}_B$ are the generators of general coordinate transformations that satisfy the commutation relations
\begin{equation}
(\sigma_j{}^i)^A{}_C(\sigma_l{}^k)^C{}_B - (\sigma_l{}^k)^A{}_C (\sigma_j{}^i)^C{}_B
= (\sigma_l{}^i)^A{}_B\,\delta^k_j - (\sigma_j{}^k)^A{}_B\,\delta^i_l.\label{comms}
\end{equation}
We immediately recognize in (\ref{comms}) the Lie algebra of the general linear group $GL(4,R)$. This fact is closely related to the standard gauge-theoretic interpretation \cite{PRs} of metric-affine gravity as the gauge theory of the general affine group $GA(4,R)$, which is a semidirect product of spacetime translation group times $GL(4,R)$.

The transformation properties (\ref{dpsiA}) determine the form of the covariant and the Lie derivative of a matter field: 
\begin{align}\label{Dpsi}
\nabla_k\psi^A &:= \partial_k\psi^A -\Gamma_{ki}{}^j\,(\sigma_j{}^i)^A{}_B\,\psi^B,\\ \label{Lpsi}
{\cal L}_\zeta\psi^A &:= \zeta^k\partial_k\psi^A + (\partial_i\zeta^j)(\sigma_j{}^i)^A{}_B\,\psi^B.
\end{align}
The commutators of these differential operators read
\begin{align}
(\nabla_k\nabla_l - \nabla_l\nabla_k)\psi^A &= -\,R_{klj}{}^i(\sigma_i{}^j)^A{}_B\psi^B - T_{kl}{}^i\nabla_i\psi^A,\label{comDDpsi}\\
 ({\cal L}_\zeta \nabla_k - \nabla_k{\cal L}_\zeta)\psi^A &= -\,({\cal L}_\zeta\Gamma_{kj}{}^i)(\sigma_i{}^j)^A{}_B\psi^B.\label{comDLpsi}
\end{align}

\subsection{Irreducible decomposition of curvature and torsion} 

In order to establish the dynamical scheme of the Poincar\'e gauge gravity in a most transparent way, and also to understand more clearly the coupling of the gravitational field to the physical sources of different physical nature, it is convenient to decompose the Poincar\'e gauge field strengths, the curvature and the torsion, into irreducible parts.

With the help of the metric $g_{ij}$ and the totally antisymmetric Levi-Civita tensor $\eta_{ijkl}$, one can construct a number of contractions of the curvature. In particular, we introduce the {\it Ricci tensor} and the {\it co-Ricci tensor} as
\begin{equation}\label{Ric}
R_{ij} := R_{kij}{}^k,\qquad \overline{R}{}^{ij} := {\frac 12}\,R_{klm}{}^i\,\eta^{klmj},
\end{equation}
respectively. By definition, the former is a parity-even object, whereas the latter is a parity-odd one. We can split (\ref{Ric}) into the skew-symmetric and symmetric pieces
\begin{equation}\label{RicAS}
R_{ij} = R_{[ij]} + R_{(ij)},\qquad \overline{R}{}^{ij} = \overline{R}{}^{[ij]} + \overline{R}{}^{(ij)},
\end{equation}
and, furthermore, extract the traceless parts from the latter
\begin{equation}\label{RicT}
\aR_{ij} := R_{(ij)} - {\frac 14} R g_{ij},\qquad \aOR{\,\,}^{ij} := \overline{R}{}^{(ij)}
- {\frac 14}\overline{R}g^{ij}.
\end{equation}
Here the curvature scalar and pseudoscalar arise naturally as the traces
\begin{equation}\label{scalars}
R = g^{ij}R_{ij} = R_{ij}{}^{ji},\qquad \overline{R} = g_{ij}\overline{R}{}^{ij} = {\frac 12}\,R_{ijkl}\,\eta^{ijkl}.
\end{equation}
With the help of a straightforward algebra we can verify that the antisymmetric Ricci parts are related via
\begin{equation}
\overline{R}{}^{[ij]} = {\frac 12}\eta^{ijkl}R_{[kl]},\label{RicAA}
\end{equation}
and it will be convenient to denote the skew-symmetric tensor as
\begin{equation}
\check{R}_{ij} := R_{[kl]}.\label{RicA}
\end{equation}

Then irreducible parts of the curvature tensor are as follows:
\begin{align}
{}^{(2)}\!R_{kl}{}^{ij} &= \aOR_m{}^{[i}\eta^{j]m}{}_{kl},\label{R2}\\
{}^{(3)}\!R_{kl}{}^{ij} &= -\,{\frac {1}{12}}\,\overline{R}\,\eta_{kl}{}^{ij},\label{R3}\\ 
{}^{(4)}\!R_{kl}{}^{ij} &= -\,2\!\!\aR_{[k}{}^{[i}\,\delta^{j]}_{l]},\label{R4}\\
{}^{(5)}\!R_{kl}{}^{ij} &= -\,2\check{R}_{[k}{}^{[i}\,\delta^{j]}_{l]},\label{R5}\\
{}^{(6)}\!R_{kl}{}^{ij} &= -\,{\frac {1}{6}}\,R\,\delta_{[k}{}^{i}\,\delta^{j}_{l]}.\label{R6}
\end{align}
In the literature, the five objects (\ref{R2})-(\ref{R6}) are known \cite{PRs} as the ``paircom'', ``pscalar'', ``ricsymf'', ``ricanti'', and ``scalar'' parts, respectively. Finally, the Weyl part is defined as
\begin{equation}
{}^{(1)}\!R_{kl}{}^{ij} = R_{kl}{}^{ij} - \sum_{I=2}^6 {}^{(I)}\!R_{kl}{}^{ij}\label{R1}
\end{equation}

Introducing the trace vector and the axial trace vector, respectively,
\begin{equation}
T_j := T_{ij}{}^i,\qquad \overline{T}{}^j = {\frac 12}T_{kli}\eta^{klij},\label{Pv}
\end{equation}
the torsion tensor decomposition reads $T_{kl}{}^{i} = {}^{(1)}T_{kl}{}^{i} + {}^{(2)}T_{kl}{}^{i} + {}^{(3)}T_{kl}{}^{i}$, with
\begin{align}
{}^{(2)}T_{kl}{}^{i} &= {\frac 23}\,\delta^i_{[k}T_{l]},\label{tT2}\\
{}^{(3)}T_{kl}{}^{i} &= -\,{\frac 13}\eta_{kl}{}^{ij}\overline{T}{}_j,\label{tT3}\\
{}^{(1)}T_{kl}{}^{i} &= T_{kl}{}^{i} - {}^{(2)}T_{kl}{}^{i} - {}^{(3)}T_{kl}{}^{i}.\label{tT1}
\end{align}

\section{General structure of Poincar\'e gauge gravity}\label{PG}

The gauging of the Poincar\'e symmetry group is well understood within the framework of a general gauge-theoretic approach which is formulated as a heuristic scheme in the Lagrange formalism in the Minkowski space of special relativity for the purpose of deriving a new interaction from a conserved Noether current associated with rigid symmetry group \cite{ORaifeartaigh:1978,Mack:1979,Chaichian:1984}. Such a new gauge interaction arises from the requirement that the rigid (global) symmetry should be extended to a local symmetry.

In contrast to the standard theories of electroweak and strong interactions, which are based on the gauging of internal symmetries, the gauge theory of the gravitational interaction is underlied by external symmetry groups of the spacetime. In the absence of gravity, the fundamental spacetime symmetry of the flat Minkowski space is its group of motions, namely, the Poincar\'e group $T_4\rtimes SO(1,3)$, the semi-direct product of the translations group $T_4$ (four parameters $\varepsilon^\alpha$) and the Lorentz group $SO(1,3)$ (six parameters $\varepsilon^{\alpha\beta} = -\,\varepsilon^{\beta\alpha}$). The corresponding Lagrange-Noether treatment of the invariance of Minkowski space under rigid (global) Poincar\'e transformations gives rise to the conservation laws of the canonical energy-momentum $\Sigma_\alpha{}^i$ and spin angular momentum  $\tau_{\a\b}{}^i = -\,\tau_{\b\a}{}^i$ currents. In relation with this, it is worthwhile to recall Wigner's classification \cite{Wigner:1939cj} of quantum mechanical systems in a Minkowski space according to {\it mass and spin.} 

An appropriate gauge-theoretic formalism that extends the approach of Yang and Mills \cite{Yang:1954} from the case of internal symmetries to the spacetime symmetry groups was developed by Utiyama, Sciama and Kibble \cite{Utiyama:1956,Sciama:1962,Kibble:1961}. Up-to-date reviews of the Poincar\'e gauge theory of gravity can be found in \cite{Hehl:2020,Obukhov:2006,Obukhov:2018}, and for more historic and technical details readers may refer to \cite{Blagojevic:2013,Ponomarev,Mielke,Erice:1979,PRs}. Here we briefly outline the most essential notions and constructions. 

\subsection{Poincar\'e gauge gravity kinematics}

Following the general Yang-Mills-Utiyama-Sciama-Kibble gauge-theoretic sche\-me, the 10-parameter Poincar\'e group $T_4\!\rtimes\!SO(1,3)$ gives rise to the 10-plet of the gauge potentials which are consistently identified with the orthonormal coframe $e_i^\alpha$ (4 potentials corresponding to the translation subgroup $T_4$) and the local connection $\Gamma_i{}^{\alpha\beta} = -\,\Gamma_{i}{}^{\beta\alpha}$ (6 potentials for the Lorentz subgroup $SO(1,3)$). The corresponding  field strengths of translations and Lorentz rotations arise as covariant ``curls''
\begin{align}
T_{ij}{}^\alpha &= \partial_i e_j{}^\a - \partial_j e_i{}^\a
+ \Gamma_{i\beta}{}^\alpha e_j{}^\b - \Gamma_{j\beta}{}^\alpha e_i{}^\b,\label{Tor}\\ \label{Cur}
R_{ij\alpha}{}^{\beta} &= \partial_i\Gamma_{j\alpha}{}^{\beta} - \partial_j\Gamma_{i\alpha}{}^{\beta} 
+ \Gamma_{i\gamma}{}^\beta\Gamma_{j\alpha}{}^{\gamma} - \Gamma_{j\gamma}{}^\beta\Gamma_{i\alpha}{}^{\gamma}.
\end{align}
Comparing these with (\ref{curv}) and (\ref{tors}), respectively, with an account of (\ref{GG2}), we immediately identify the Poincar\'e gauge field strengths (\ref{Tor}) and (\ref{Cur}) with the torsion $T_{ij}{}^\a = e^\alpha_k T_{ij}{}^k$ and the curvature $R_{ij\alpha}{}^{\beta} = e_\alpha^k e^\beta_k R_{ijk}{}^l$ of the Riemann-Cartan geometry on the spacetime manifold.

In accordance with the heuristic gauging scheme, the gravitational spin-connection interaction is derived from the rigid Lorentz symmetry of a matter field $\psi^A$ which belongs to a representation of the Lorentz group with the generators $(\rho_{\alpha\beta})^A{}_B$:
\begin{equation}\label{infP}
\delta\psi^A = - \,{\frac 12}\varepsilon^{\alpha\beta}(\rho_{\alpha\beta})^A{}_B\,\psi^B.
\end{equation}
When the transformation is extended to the local one with infinitesimal parameters $\varepsilon^{\alpha\beta} = \varepsilon^{\alpha\beta}(x)$, the covariant derivative is introduced by
\begin{equation}\label{Dp}
D_i\psi^A = \partial_i\psi^A - {\frac 12}\Gamma_i{}^{\alpha\beta}(\rho_{\alpha\beta})^A{}_B\,\psi^B.
\end{equation}

The Poincar\'e gauge field strengths satisfy the Bianchi identities, cf. (\ref{Dcurv}) and (\ref{Dtors}):
\begin{align}
D_{[k}T_{ij]}{}^\alpha &= e_{[k}{}^\beta R_{ij]\beta}{}^\alpha\,,\label{Bianchi2c}\\
D_{[k}R_{ij]}{}^{\alpha\beta} &= 0.\label{Bianchi1c}
\end{align}

Note that the rigid Lie algebra of the Poincar\'e group is extended to a so-called deformed, soft, or local ``Lie algebra'' ($D_\a$ and $\rho_{\a\b} = -\,\rho_{\b\a}$ generate translations and Lorentz transformations, respectively):
\begin{equation}\label{soft}
\left.
\begin{aligned}
\null  [D_\a,D_\b] &= -\,T_{\alpha\beta}{}^{\gamma}
  D_{\gamma}+R_{\alpha\beta}{}^{\gamma\delta}
  \rho_{\delta\gamma}\  \\
  [\rho_{\alpha\beta},D_\gamma] &= -\,g_{\gamma\alpha}D_{\beta} + g_{\gamma\beta}D_{\alpha}\ \\
  [\rho_{\alpha\beta},\rho_{\mu\nu}] &= -\,g_{\alpha\mu}\rho_{\beta\nu} + g_{\alpha\nu}\rho_{\beta\mu}
  + g_{\beta\mu}\rho_{\alpha\nu} - g_{\beta\nu}\rho_{\alpha\mu}\
\end{aligned}
\right\}.
\end{equation}
The rigid Lie algebra of Minkowski space is recovered for $T_{\a\b}{}^\g=0$ and $R_{\a\b}{}^{\g\d}=0$, when $D_\a\rightarrow \partial_a$ in Cartesian coordinates, for details see \cite{Erice:1979}.

\subsection{Poincar\'e gauge gravity dynamics: Yang-Mills type models}

Assuming the standard minimal coupling, the total Lagrangian of interacting gravitational and matter fields reads
\begin{equation}\label{Ltot}
L = V(g_{ij}, R_{ijk}{}^l, T_{ki}{}^j) + L_{\rm mat}(g_{ij}, \psi^A, \nabla_i\psi^A).
\end{equation}
In general, the gravitational Lagrangian $V$ is constructed as a diffeomorphism invariant function of the curvature and torsion. The matter Lagrangian $L_{\rm mat}$ depends on the matter fields $\psi^A$ and their covariant derivatives $\nabla_i\psi^A$.

Let us now specialize to the general quadratic model with the Lagrangian that contains all possible quadratic invariants of the torsion and the curvature:
\begin{align}
V &= -\,{\frac {1}{2\kappa c}}\biggl\{a_0R + \overline{a}{}_0\overline{R} + 2\lambda_0 
+\,{\frac 12}\sum_{I=1}^3\Bigl[a_I\,{}^{(I)}\!T^{kl}{}_i\,T_{kl}{}^i
- {\frac {\overline{a}_I}2}\,{}^{(I)}\!T_{mni}\,T_{kl}{}^i\,\eta^{mnkl}\Bigr]\nonumber\\
&\hspace{12mm} +\,{\frac {\ell_\rho^2}{2}}\sum_{I=1}^6 \Bigl[b_I\,{}^{(I)}\!R^{kl}{}_{ij}\,R_{kl}{}^{ij}
- {\frac {\overline{b}_I}2}\,{}^{(I)}\!R_{mnij}\,R_{kl}{}^{ij}\,\eta^{mnkl}\Bigr]\biggr\}.\label{LRT}
\end{align}
Here $\kappa = {\frac {8\pi G}{c^3}}$ is Einstein's gravitational constant with the dimension of $[\kappa] = $s\,kg$^{-1}$. $G = 6.67\times 10^{-11}$ m$^3$\,kg$^{-1}$\,s$^{-2}$ is Newton's gravitational constant. The speed of light $c = 2.9\times 10^8$ m/s. 

Besides the linear ``Hilbert type'' part characterized by $a_0$ and $\overline{a}_0$, the Lagrangian (\ref{LRT}) contains several additional coupling constants which fix the ``Yang-Mills type'' part: $a_1, a_2, a_3$, $\overline{a}_1, \overline{a}_2, \overline{a}_3$, $b_1, \cdots, b_6$, $\overline{b}_1, \cdots, \overline{b}_6$, and $\ell_\rho^2$. The latter has the dimension $[{\ell_\rho^2}]=\,$[area] so that $[{\ell_\rho^2}/{\kappa c}] = [\hbar]$, whereas $a_I$, $\overline{a}_I$, $b_I$ and $\overline{b}_I$ are dimensionless. Moreover, not all of these constants are independent: in the parity-odd sector we take $\overline{b}_2 = \overline{b}_4$ and $\overline{b}_3 = \overline{b}_6$ because the two pairs of terms in (\ref{LRT}) are the same: 
\begin{align}
{}^{(2)}\!R_{mnij}\,R_{kl}{}^{ij}\,\eta^{mnkl} = {}^{(4)}\!R_{mnij}\,R_{kl}{}^{ij}\,\eta^{mnkl} 
= {}^{(2)}\!R_{mnij}\,{}^{(4)}\!R_{kl}{}^{ij}\,\eta^{mnkl},\label{R24} \\ 
{}^{(3)}\!R_{mnij}\,R_{kl}{}^{ij}\,\eta^{mnkl} = {}^{(6)}\!R_{mnij}\,R_{kl}{}^{ij}\,\eta^{mnkl} 
= {}^{(3)}\!R_{mnij}\,{}^{(6)}\!R_{kl}{}^{ij}\,\eta^{mnkl},\label{R36}
\end{align}
whereas ${}^{(1)}\!R_{mnij}\,R_{kl}{}^{ij}\,\eta^{mnkl} = {}^{(1)}\!R_{mnij}\,{}^{(1)}\!R_{kl}{}^{ij}\,\eta^{mnkl}$ and ${}^{(5)}\!R_{mnij}\,R_{kl}{}^{ij}\,\eta^{mnkl} = {}^{(5)}\!R_{mnij}\,{}^{(5)}\!R_{kl}{}^{ij}\,\eta^{mnkl}$. One can prove these relations directly from the definitions (\ref{R2})-(\ref{R1}). Similarly (\ref{tT2})-(\ref{tT1}) yield
\begin{equation}
{}^{(2)}\!T_{mni}\,T_{kl}{}^i\,\eta^{mnkl} = {}^{(3)}\!T_{mni}\,T_{kl}{}^i\,\eta^{mnkl} 
= {}^{(2)}\!T_{mni}\,{}^{(3)}\!T_{kl}{}^i\,\eta^{mnkl},\label{T23}
\end{equation}
which leads to a constraint  $\overline{a}_2 = \overline{a}_3$.

For completeness, we included the cosmological constant $\lambda_0$ with the dimension of the inverse area, $[\lambda_0] = [\ell^{-2}]$. In the special case $a_0 = \overline{a}_0 = 0$ the purely quadratic model is obtained without the Hilbert-Einstein linear term in the Lagrangian. 

In the literature, the quadratic Poincar\'e gravity theories are often formulated in terms of the standard tensor objects which are not decomposed into irreducible parts. In order to be able to compare (\ref{LRT}) to the models studied in the literature,  let us rewrite the Lagrangian $V$ explicitly:
\begin{align}
V &= -\,{\frac 1{2\kappa c}}\Bigl\{a_0 R + \overline{a}{}_0\overline{R} + 2\lambda_0\nonumber\\ 
& \hspace{12mm} +\,\alpha_1\,T_{kl}{}^i\,T^{kl}{}_i + \alpha_2\,T_i\,T^i + \alpha_3\,T_{kl}{}^i\,T_i{}^{kl}\nonumber\\
& \hspace{12mm} +\,\overline{\alpha}_1\,\eta^{klmn}\,T_{kli}\,T_{mn}{}^i + \overline{\alpha}_2\,\eta^{klmn}
\,T_{klm}\,T_n\nonumber\\
&\hspace{5mm} +\,{\ell_\rho^2}\Big(\beta_1\,R_{ijkl}R^{ijkl} + \beta_2\,R_{ijkl}R^{ikjl} 
+ \beta_3\,R_{ijkl}R^{klij}\nonumber\\
&\hspace{12mm} +\,\beta_4\,R_{ij}R^{ij} + \beta_5\,R_{ij}R^{ji} + \beta_6\,R^2\nonumber\\
&\hspace{12mm} +\,\overline{\beta}_1\,\eta^{klmn}\,R_{klij}R_{mn}{}^{ij}
+ \overline{\beta}_2\,\eta^{klmn}\,R_{kl}R_{mn}\nonumber\\
&\hspace{12mm} +\,\overline{\beta}_3\,\eta^{klmn}\,R_{klm}{}^i\,R_{ni}
+ \overline{\beta}_4\,\eta^{klmn}\,R_{klmn}\,R\Big)\Bigr\}.\label{lagrT}
\end{align}
Using the definitions of the irreducible torsion (\ref{tT2})-(\ref{tT1}) and curvature 
(\ref{R2})-(\ref{R1}) parts, we find the relation between the coupling constants:
\begin{align}
a_1 &= 2\alpha_1 - \alpha_3,\quad a_2 = 2\alpha_1 + 3\alpha_2 - \alpha_3,
\quad a_3 = 2\alpha_1 + 2\alpha_3,\label{ccc}\\
\overline{a}_1 &= -\,4\overline{\alpha}_1,\quad \overline{a}_2 = \overline{a}_3
= -\,4\overline{\alpha}_1 - 3\overline{\alpha}_2,\\
b_1 &= 2\beta_1 + \beta_2 + 2\beta_3,\\
b_2 &= 2\beta_1 - 2\beta_3,\\
b_3 &= 2\beta_1 - 2\beta_2 + 2\beta_3,\\
b_4 &= 2\beta_1 +  \beta_2 + 2\beta_3 +  \beta_4 +  \beta_5,\\
b_5 &= 2\beta_1 - 2\beta_3 + \beta_4 - \beta_5,\\
b_6 &= 2\beta_1 +  \beta_2 + 2\beta_3 + 3\beta_4 + 3\beta_5 + 12\beta_6,\\
\overline{b}_1 &= -\,4\overline{\beta}_1,\quad \overline{b}_2 = \overline{b}_4
= -\,4\overline{\beta}_1 + \overline{\beta}_3,\\
\overline{b}_5 &= -\,4\overline{\beta}_1 - 2\overline{\beta}_2 + 2\overline{\beta}_3,
\quad \overline{b}_3 = \overline{b}_6 = -\,4\overline{\beta}_1 + 3\overline{\beta}_3
+ 12\overline{\beta}_4.
\end{align}
The inverse of (\ref{ccc}) reads
\begin{equation}
\alpha_1 = {\frac {2a_1 + a_3} 6},\quad \alpha_2 = {\frac {a_2 - a_1} 3},
\quad \alpha_3 = {\frac {a_3 - a_1} 3}.
\end{equation}
Not all terms in the Lagrangian (\ref{lagrT}) are independent, since the expressions
\begin{align}
V_{\rm GB} &= -\,{\frac 14}\,\eta^{klmn}\eta_{ijpq}R_{kl}{}^{ij}R_{mn}{}^{pq}
= R_{klij}R^{ijkl} - 4 R_{ij}R^{ji} + R^2,\label{GB}\\
V_{\rm PC} &= {\frac 12}\,\eta^{klmn}R_{klij}R_{mn}{}^{ij},\label{PC}\\
V_{\rm NY} &= \overline{R} + {\frac 12}\,\eta^{klmn}T_{kli}T_{mn}{}^i,\label{NY}
\end{align}
are the total divergences. Integrating these scalar quantities (with appropriate normalization factors) over the spacetime manifold, one obtains the topological invariants \cite{Eguchi:1980,Hehl:1991,Chandia:1998} known as the Euler, Pontryagin (or Chern), and Nieh-Yan characteristics, respectively. Therefore, some of the constants $\beta_3$, $\beta_5$, $\beta_6$, $\overline{\beta}_1$, $\overline{a}_0$, and $\overline{\alpha}_1$, may be eliminated (same applies to the set of constants $b_I$, $\overline{b}_I$, $\overline{a}_J$). However, here this possibility is not used.

The Poincar\'e gauge gravity field equations arise from the variation of the total action with respect to the coframe and connection. They read explicitly
\begin{align}
a_0\Bigl(R_i{}^j - {\frac 12}R\delta_i^j\Bigr) - \overline{a}_0\overline{R}_i{}^j - \lambda_0\delta_i^j
&\nonumber\\ +\,{\stackrel {(T)}q}{}_i{}^j + \ell_\rho^2{\stackrel {(R)}q}{}_i{}^j 
- \Bigl[(\nabla_l - T_l)h^{jl}{}_i + {\frac 12}T_{mn}{}^jh^{mn}{}_i\Bigr]
&= \kappa \Sigma_i{}^j,\label{PG1}\\
a_0\left(T_{ij}{}^k + 2T_{[i}\delta_{j]}^k\right) - {\frac {\overline{a}_0}{2}}\,\eta_{ij}{}^{mn}
\left(T_{mn}{}^k + 2T_{[m}\delta_{n]}^k\right) - 2h^k{}_{[ij]}\nonumber\\
- 2\ell_\rho^2\Bigl[(\nabla_l - T_l)h^{kl}{}_{ij} + {\frac 12}T_{mn}{}^kh^{mn}{}_{ij}\Bigr]
&= \kappa c\tau_{ij}{}^k.\label{PG2}  
\end{align}
Here the gravitational momenta are described by
\begin{align}
h^{ij}{}_k &= \sum_{I=1}^3\,a_I\,{}^{(I)}\!T^{ij}{}_k - {\frac 12}\eta^{ij}{}_{mn}
\sum_{I=1}^3\,\overline{a}_I\,{}^{(I)}\!T^{mn}{}_k,\label{HT}\\
h^{ij}{}_{kl} &= \sum_{I=1}^6\,b_I\,{}^{(I)}\!R^{ij}{}_{kl} - {\frac 12}\eta^{ij}{}_{mn}
\sum_{I=1}^6\,\overline{b}_I\,{}^{(I)}\!R^{mn}{}_{kl},\label{HR}
\end{align}
from which we construct the two objects quadratic in the torsion and the curvature
\begin{align}\label{qT}
{\stackrel {(T)}q}{}_i{}^j &= T_{in}{}^kh^{jn}{}_k - {\frac 14}\delta_i^j\,T_{mn}{}^kh^{mn}{}_k,\\
{\stackrel {(R)}q}{}_i{}^j &= R_{in}{}^{kl}h^{jn}{}_{kl} - {\frac 14}\delta_i^j\,R_{mn}{}^{kl}
h^{mn}{}_{kl}.\label{qR}
\end{align}

\subsubsection{Example: Matter with spin}

In order to give an explicit example of a typical physical matter source of the gravitational field in PG theory, we recall the classical model of spinning fluid \cite{itm:188}. This model was first worked out by Weyssenhoff and Raabe \cite{Weyssenhoff:1947} as a direct development of the ideas of Cosserats \cite{Cosserat} who proposed to describe the microstructure properties of a medium by attaching a rigid material frame to every element of a continuum. Using the variational principle for the spinning fluid \cite{Obukhov:1987}, one derives the canonical energy-momentum and spin tensors:
\begin{align}
\Sigma_j{}^i &= \,u^i{\mathcal P}_j - p\Bigl(\delta_j^i - {\frac {1}{c^2}}u_ju^i\Bigr),\label{hypS}\\
\tau_{ij}{}^k &= \,u^k{\mathcal S}_{ij},\label{hypD}
\end{align}
where $u^i$ is the 4-velocity of the fluid and $p$ is the pressure. Fluid elements are characterized by their microstructural properties: the energy density $\varepsilon$, the intrinsic spin density ${\mathcal S}_{ij} = -\,{\mathcal S}_{ji}$ (subject to the Frenkel supplementary condition ${\mathcal S}_{ij}u^j = 0$), and the momentum density 
\begin{equation}
{\mathcal P}_i = {\frac {1}{c^2}}\left[\varepsilon u_i
+ u^j(\nabla_k - T_k)({\mathcal S}_{ij}u^k)\right].\label{sP}
\end{equation}

\subsection{Parity-even model: particle spectrum}

To streamline the subsequent discussion, we now specialize to the {\it purely parity-even model} and hence assume $\overline{a}_0 = 0$, $\overline{a}_I = 0$, and $\overline{b}_I = 0$ in the rest of the paper.

The number of graviton modes in PG theory (that mediate the gravitational interaction) is much larger than in Einstein's theory. The analysis of the particle spectrum for the quadratic model (\ref{LRT}) reveals \cite{itm:134,itm:135,Obukhov:1989} that the dynamics of graviton modes in different $J^P$ (spin$^{\rm parity}$) sectors is determined by the following combinations of the coupling constants:
\begin{eqnarray}
&&\left.\begin{split}
2^+:\,\Lambda_1 \,&=\, b_1 + b_4,\\
2^-:\,\Lambda_2 \,&=\, b_1 + b_2,\\
0^+:\,\Lambda_3 \,&=\, b_4 + b_6,\\
0^-:\,\Lambda_4 \,&=\, b_2 + b_3,\\
1^+:\,\Lambda_5 \,&=\, b_2 + b_5,\\
1^-:\,\Lambda_6 \,&=\, b_4 + b_5,
\end{split}\quad\right\}\label{LL}\\
&&\left.\begin{split}
\mu_1 \,&=\, -\,a_0 + 2a_3,\\
\mu_2 \,&=\, -\,2a_0 + a_2,\\
\mu_3 \,&=\, -\,a_0 - a_1,
\end{split}\hspace{7mm}\right\}\label{mumu}
\end{eqnarray}
which specify the corresponding kinetic and mass terms for these modes. The mapping (\ref{LL}) between the two sets of coupling constants $b_I$ and $\Lambda_I$ is not one-to-one. Namely, whereas $b_I = 0$ yields $\Lambda_I = 0$, inverse is not true, and from $\Lambda_I = 0$ we find $b_1 = - b_2 = b_3 = - b_4 = b_5 = b_6$, which brings the curvature square part of the gravitational Lagrangian (\ref{lagrT}) to $b_1V_{\rm GB}$, cf. (\ref{GB}). 

The general analysis of the particle spectrum with both parity-even and parity-odd sectors included can be found in the recent papers \cite{Karananas:2015,Blagojevic:2018}.

\section{Physical consequences of Poincar\'e gauge gravity}\label{sec6.3}

In this introductory review, we do not aim to deal with all the aspects of PG theory. Instead, we will focus on the issues of correspondence of PG and GR for the whole class of models (\ref{LRT}) and then will discuss in some more detail several specific models that were studied in the literature. 

\subsection{Correspondence with GR: Torsionless solutions}

Let us discuss the correspondence of the vacuum field equations (\ref{PG1})-(\ref{PG2}) for the Yang-Mills type quadratic Lagrangian (\ref{LRT}) and Einstein's general relativity under the assumption of {\it vanishing torsion}, $T_{kl}{}^i = 0$. Then we find $h^{ij}{}_k = 0$, hence ${\stackrel {(T)}q}{}_i{}^j = 0$, whereas $\overline{R}_i{}^j = 0$ and the curvature has only three nontrivial parts (\ref{R4}), (\ref{R6}), (\ref{R1}). A direct computation yields
\begin{equation}\label{qRnot}
{\stackrel {(R)}q}{}_i{}^j = \Lambda_1 {}^{(1)}\!R_{ikl}{}^j\!\aR^{kl} + {\frac {\Lambda_3}{6}}R\!\aR_i{}^j,
\end{equation}
and in vacuum (when $\Sigma_i{}^j = 0$ and $\tau_{ij}{}^k = 0$) the field equations (\ref{PG1})-(\ref{PG2}) reduce to
\begin{align}\label{PG1not}
a_0\Bigl(R_i{}^j - {\frac 12}R\delta_i^j\Bigr) - \lambda_0\delta_i^j
+ \ell_\rho^2\Lambda_1 {}^{(1)}\!R_{ikl}{}^j\!\aR^{kl} + {\frac {\ell_\rho^2\Lambda_3}{6}}R\!\aR_i{}^j &= 0,\\
\Lambda_1\nabla_l{}^{(1)}\!R^{kl}{}_{ij} + {\frac {\Lambda_3}{6}}\,\nabla_{[i}R\,\delta_{j]}^k &= 0.\label{PG2not}
\end{align}
As we see, only two coupling constants $\Lambda_1$ and $\Lambda_3$ enter the field equations. This is consistent with the fact that they essentially determine the structure of the effective Lagrangian obtained from (\ref{lagrT}) for the vanishing torsion, $T_{kl}{}^i = 0$:
\begin{align}
V &= -\,{\frac {1}{2\kappa c}}\Bigl( a_0 R + 2\lambda_0 + {\frac {\ell_\rho^2}{12}}\left\{
(4\Lambda_1 - \Lambda_3)\,R_{ijkl}R^{ijkl} + 4(\Lambda_3 - \Lambda_1)\,R_{ij}R^{ij}\right\}\nonumber \\
&\hspace{2cm} +\,\ell_\rho^2\beta_6\left\{R_{ijkl}R^{klij} - 4R_{ij}R^{ji} + R^2\right\}\Bigr).
\end{align}
The last line does not contribute to the field equations, being a total divergence of the Euler (Gauss-Bonnet) topological term (\ref{GB}). When $\Lambda_1 = 0$ and $\Lambda_3 = 0$, the field equations reduce to Einstein's equation with the cosmological term. 

Contracting (\ref{PG1not}), we find
\begin{equation}
a_0R = -\,4\lambda_0,\label{PG1T}
\end{equation}
and, provided $a_0 \neq 0$, the system (\ref{PG1not})-(\ref{PG2not}) is recast into
\begin{align}\label{PG1not2}
a_0^{\rm eff}\!\aR_i{}^j + \ell_\rho^2\Lambda_1 {}^{(1)}\!R_{ikl}{}^j\!\aR^{kl} &= 0,\\
\Lambda_1\left(\nabla_i\!\aR_j{}^k - \nabla_j\!\aR_i{}^k\right) &= 0.\label{PG2not2}
\end{align}
where the effective constant is introduced by
\begin{equation}\label{a0eff}
a_0^{\rm eff} = a_0 + {\frac {\ell_\rho^2\Lambda_3R}{6}}
= a_0 - {\frac {2\ell_\rho^2\Lambda_3\lambda_0}{3a_0}}.
\end{equation}    
The last equation (\ref{PG2not2}) follows from the Bianchi identity (\ref{Dcurv}).

It is obvious that the vacuum Einstein spaces \cite{Petrov} with a cosmological term (\ref{PG1T})
\begin{equation}\label{eq:6.25}
\aR_{ij} = 0,\qquad {\rm i.e.,}\qquad R_{ij} = {\frac 14}\,R\,g_{ij}\,,
\end{equation}
are solutions of the system (\ref{PG1not2})-(\ref{PG2not2}). The questions is: are there other solutions, or (\ref{eq:6.25}) represents the unique solution? When $a_0 \neq 0$, there are several situations depending on the values of $\Lambda_1$ and $\Lambda_3$.

When $\Lambda_1 = 0$, the system (\ref{PG1not})-(\ref{PG2not}) reduces to
\begin{equation}\label{eq:6.26}
a_0^{\rm eff}\!\aR_{ij} = 0.
\end{equation}
Then we have one of the two possibilities. If $a_0^{\rm eff} \neq 0$ the system (\ref{eq:6.26}) coincides with Einstein's field equations (\ref{eq:6.25}). In the special case $a_0^{\rm eff} = 0$, equations (\ref{PG1not2})-(\ref{PG2not2}) are fulfilled identically, and the solutions are arbitrary spaces which satisfy $a_0R = -\,4\lambda_0$.

If $\Lambda_1 \neq 0$, we introduce
\begin{equation}\label{xill}
\xi := {\frac {a_0^{\rm eff}}{\ell_\rho^2\Lambda_1}}
\end{equation}
and the system (\ref{PG1not2})-(\ref{PG2not2}) is then recast into
\begin{align}
&{}^{(1)}\!R_{iklj}\!\aR^{kl} = -\,\xi\!\aR^{kl},\label{PG1not3}\\
&\nabla_i\!\aR_{jk} - \nabla_j\!\aR_{ik} = 0.\label{PG2not3}
\end{align}

One can prove (for technical details and the references, see \cite{Obukhov:1989}) that the only solutions of the system (\ref{PG1not2})-(\ref{PG2not2}) are Einstein spaces (\ref{eq:6.25}), provided
\begin{equation}\label{exc}
\xi \neq \left\{0,\ -\,{\frac {2\lambda_0}{3a_0}},\ {\frac {4\lambda_0}{3a_0}}\right\}.
\end{equation}
When $\xi$ takes one of the exceptional values listed in (\ref{exc}), the system (\ref{PG1not2})-(\ref{PG2not2}) admits solutions with $\aR_{ij}\neq 0$ which are not Einstein spaces \cite{itm:189}.

For completeness, let us mention that similar conclusions can be derived for the purely quadratic model with $a_0 = 0$. In this case the cosmological constant should vanish $\lambda_0 = 0$, and we find $\xi = \Lambda_3R/6\Lambda_1$.

\subsection{Correspondence with GR: Birkhoff theorem}

Let us now return to the general case, and discuss the correspondence of PG and GR without assuming vanishing torsion. In view of the fact that the fundamental gravitational experiments in our Solar system are perfectly consistent with the Schwarzschild geometry, a natural question arises: to which extent the solutions of the field equations in PG theory may deviate from the Schwarzschild spacetime? 

Quite generally, spherically symmetric solutions are of particular interest in all field-theoretic models. In Einstein's general relativity theory, the Schwarzschild metric is a unique solution of the gravitational field equations under the assumption of a spherical symmetry of the spacetime geometry and matter source distribution. This remarkable result is known as the Birkhoff theorem.

In order to discuss the validity of the {\it generalized Birkhoff theorem} in Poincar\'e gauge gravity, we need to clarify how the spherical symmetry is described for the gravitational gauge fields \cite{MH1,MH2,birk}. Following Sec.~\ref{killing_sec}, in the local coordinate system $x^i = (t, r, \theta, \varphi)$, the most general spherically symmetric spacetime interval 
\begin{equation}
ds^2 = A^2dt^2 - B^2dr^2 - C^2(d\theta^2 + \sin^2\theta d\varphi^2)\label{Sds}
\end{equation}
depends on the three arbitrary functions $A = A(t,r)$, $B = B(t,r)$, $C = C(t,r)$. An $SO(3)$ rotation motion of the manifold $M$ is generated by three vector fields
\begin{equation}
\xi_{\{x\}}^i = \begin{footnotesize}\left(\begin{array}{c}
0\\ 0\\ \sin\varphi \\ \cos\varphi\cot\theta\end{array}\right)\end{footnotesize},\qquad
\xi_{\{y\}}^i = \begin{footnotesize}\left(\begin{array}{c}
0\\ 0\\ -\cos\varphi \\ \sin\varphi\cot\theta\end{array}\right)\end{footnotesize},\qquad
\xi_{\{z\}}^i = \begin{footnotesize}\left(\begin{array}{r}
0\\ 0\\ 0\\ -1\end{array}\right)\end{footnotesize},\label{Sgen}
\end{equation}
and the spherical invariance is manifest in the vanishing Lie derivative of the metric ${\cal L}_\xi\,g_{ij} = 0$ under the action $\xi: SO(3)\times M \rightarrow M$.

In the framework of Poincar\'e gauge gravity, the general spherically symmetric configuration for the gravitational gauge field potentials ($e^\alpha_i, \Gamma_{i\beta}{}^\alpha$) reads
\begin{equation}
e^\alpha_i = \begin{footnotesize}\left(\begin{array}{cccc}
\,A\,\, & \,\,0\,\, & \,\,0\,\, & \,\,0\, \\ \,0\,\, & \,\,B\,\, & \,\,0\,\, & \,\,0\,\\
\,0\,\, & \,\,0\,\, & \,\,C\,\, & \,\,0\, \\
\,0\,\, & \,\,0\,\, & \,\,0\,\, & \,\,C\sin\theta\,\end{array}\right)\end{footnotesize},\label{Scof}
\end{equation}
for the coframe, and for the connection (using an obvious matrix notation):
\begin{alignat}{2}
&\Gamma_{0\beta}{}^\alpha = \begin{footnotesize}\left(\begin{array}{cccc}
\,0\,\, & \,\,f\,\, & \,\,0\,\, & \,\,0\, \\ \,f\,\, & \,\,0\,\, & \,\,0\,\, & \,\,0\,\\
\,0\,\, & \,\,0\,\, & \,\,0\,\, & \,\,\overline{f}\, \\
\,0\,\, & \,\,0\,\, & \,\,-\overline{f}\,\, & \,\,0\,\end{array}\right)\end{footnotesize},  \quad &
&\Gamma_{1\beta}{}^\alpha = \begin{footnotesize}\left(\begin{array}{cccc}
\,0\,\, & \,\,g\,\, & \,\,0\,\, & \,\,0\, \\ \,g\,\, & \,\,0\,\, & \,\,0\,\, & \,\,0\,\\
\,0\,\, & \,\,0\,\, & \,\,0\,\, & \,\,\overline{g}\, \\
\,0\,\, & \,\,0\,\, & \,\,-\overline{g}\,\, & \,\,0\,\end{array}\right)\end{footnotesize},\label{Sgam1}\\
&\Gamma_{2\beta}{}^\alpha = \begin{footnotesize}\left(\begin{array}{cccc}
\,0\,\, & \,\,0\,\, & \,\,p\,\, &\,\,\overline{q}\, \\
\,0\,\, & \,\,0\,\, & \,\,q\,\, & \,\,-\overline{p}\,\\
\,p\,\, & \,\,-q\,\, & \,\,0\,\, & \,\,0\, \\ 
\,\overline{q}\,\, & \,\,\overline{p}\,\, & \,\,0\,\, & \,\,0\,\end{array}\right)\end{footnotesize}, \quad &
&\Gamma_{3\beta}{}^\alpha = \begin{footnotesize}\sin\theta\,\left(\begin{array}{cccc}
\,0\,\, & \,\,0\,\, & \,\,-\overline{q}\,\, &\,\,p\, \\
\,0\,\, & \,\,0\,\, & \,\,\overline{p}\,\, & \,\,q\,\\
\,-\overline{q}\,\, & \,\,-\overline{p}\,\, & \,\,0\,\, & \,\,-\cot\theta\, \\ 
\,p\,\, & \,\,-q\,\, & \,\,\cot\theta\,\, & \,\,0\,\end{array}\right)\end{footnotesize}.\label{Sgam2}
\end{alignat}
This increases the number of arbitrary functions by eight more variables: $f = f(t,r)$, $g = g(t,r)$, $p = p(t,r)$, $q = q(t,r)$, and $\overline{f} = \overline{f}(t,r)$, $\overline{g} = \overline{g}(t,r)$, $\overline{p} = \overline{p}(t,r)$, $\overline{q} = \overline{q}(t,r)$. 

The gauge potentials (\ref{Scof})-(\ref{Sgam2}) are constructed in full agreement with the analysis in Sec.~\ref{killing_sec}, and they satisfy the generalized invariance conditions (\ref{LRC2})-(\ref{LRC3}):
\begin{equation}
{\cal L}_\xi\,e^\alpha_i = {\stackrel {\xi}{\lambda}}{}_\beta{}^\alpha\,e^\beta_i,\qquad
{\cal L}_\xi\,\Gamma_{i\beta}{}^\alpha = -\,D_i{\stackrel {\xi}{\lambda}}{}_\beta{}^\alpha,\label{LieD}
\end{equation}
where the Lorentz algebra-valued ${\stackrel {\xi}{\lambda}}{}^{\alpha\beta} = -\,{\stackrel {\xi}{\lambda}}{}^{\beta\alpha}$ parameter is determined by vector fields which generate symmetries. For the rotation symmetry generators (\ref{Sgen}) we have explicitly
\begin{equation*}
{\stackrel {\xi_{\{x\}}}{\lambda}}{}_\beta{}^\alpha = {\frac {\cos\varphi}{\sin\theta}}
\!\begin{footnotesize}\left(\begin{array}{cccc}
\,0\,\, & \,\,0\,\, & \,\,0\,\, & \,\,0\, \\ \,0\,\, & \,\,0\,\, & \,\,0\,\, & \,\,0\,\\
\,0\,\, & \,\,0\,\, & \,\,0\,\, & \,\,1\, \\ \,0\,\, & \,\,0\,\, & \,\,-1\,\, & \,\,0\,
\end{array}\right)\end{footnotesize},\quad
{\stackrel {\xi_{\{y\}}}{\lambda}}{}_\beta{}^\alpha = {\frac {\sin\varphi}{\sin\theta}}
\!\begin{footnotesize}\left(\begin{array}{cccc}
\,0\,\, & \,\,0\,\, & \,\,0\,\, & \,\,0\, \\ \,0\,\, & \,\,0\,\, & \,\,0\,\, & \,\,0\,\\
\,0\,\, & \,\,0\,\, & \,\,0\,\, & \,\,1\, \\ \,0\,\, & \,\,0\,\, & \,\,-1\,\, & \,\,0\,
\end{array}\right)\end{footnotesize},\quad
{\stackrel {\xi_{\{z\}}}{\lambda}}{}_\beta{}^\alpha = 0.
\end{equation*}

The general spherically symmetric configuration is only invariant under the group of proper rotations $SO(3)$, however, it is not invariant under spatial reflections when the parity-odd functions $\overline{f}$, $\overline{g}$, $\overline{p}$, $\overline{q}$ are nonzero. By demanding also the invariance under reflections, one extends the symmetry group from $SO(3)$ to the full rotation group $O(3)$, and such an extension imposes an additional condition on field configurations, which forbids parity-odd variables: $\overline{f} = \overline{g} = \overline{p} = \overline{q} = 0$.

The generalized Birkhoff theorem in the Poincar\'e gauge gravity is much more nontrivial \cite{Rama,RauchGRG:1982,Rauch:1981,RauchPRD:1982,Neville:1978,Neville:1980,Cruz} than its Riemannian analogue in GR, since besides the metric (coframe) variables $A, B, C$ there are additional connection variables $f, g, p, q$ and $\overline{f}, \overline{g}, \overline{p}, \overline{q}$, and the torsion is not assumed to be zero. To prove the generalized Birkhoff theorem, one needs to plug the spherically symmetric ansatz (\ref{Scof})-(\ref{Sgam2}) into the field equations (\ref{PG1})-(\ref{PG2}) and then establish the conditions under which these field equations admit only solutions with the vanishing torsion and the Schwarzschild metric. There are different types of conditions: some of them may restrict the coupling constants (hence, refine the structure of the Lagrangian $V$), other conditions may impose constraints on the geometric properties of spacetime. Among the latter are: an asymptotic flatness condition which requires that in the limit of $r\rightarrow\infty$ the metric approaches the Minkowski line element, i.e. $A \rightarrow 1$, $B \rightarrow 1$, $C \rightarrow r$, or an assumption of the vanishing scalar curvature.

There are two versions of the generalized Birkhoff theorem in the Poincar\'e gauge gravity \cite{Obukhov:1989,birk}: the strong ({\it SB}) and the weak ({\it WB}) ones. The {\it strong $SO(3)$ theorem} reads: under the assumption of the spherical symmetry in the sense of invariance under the proper rotation $SO(3)$ group, the Schwarzschild (Kottler, in general, when the cosmological constant is nontrivial) spacetime with zero torsion is a unique solution of the vacuum field equations. This result holds for the four families in the class of quadratic models (\ref{LRT}): 
\begin{description}
\item[{\it (SB1)}] No curvature square terms, $b_I = 0$ (thus $\Lambda_I = 0$), provided $\mu_1\mu_2\mu_3 \neq 0$.

\item[{\it (SB2)}] $\Lambda_1 = \Lambda_2 = \Lambda_5 = \Lambda_6 = 0$, $\Lambda_3 \neq 0, \Lambda_4 \neq 0$, provided the scalar curvature $R$ is constant, and $\mu^{\rm eff}_1\mu^{\rm eff}_2\mu^{\rm eff}_3 \neq 0$, where $\mu^{\rm eff}_I$ are obtained from (\ref{mumu}) by replacing $a_0$ with an effective coupling constant (\ref{a0eff}). 

\item[{\it (SB3)}] $\Lambda_I = 0$, provided $\mu_1\mu_2\mu_3 \neq 0$. This case is close  to ({\it SB1}) but not quite equivalent, see the remark below (\ref{mumu}).

\item[{\it (SB4)}] No torsion square terms, $a_1 = a_2 = a_3 = 0$, but $a_0 \neq 0$ and $\Lambda_1 = \Lambda_2 = \Lambda_4 = \Lambda_5 = \Lambda_6 = 0$, $\Lambda_3 \neq 0$, provided $a^{\rm eff}_0 \neq 0$.
\end{description}

It is important to notice that in the latter case the gravitational field equations yield for the curvature scalar $a_0R = -\,4\lambda_0$, and thereby the condition $a^{\rm eff}_0 \neq 0$ actually means that the constant $\xi \neq 0$, cf. (\ref{exc}).

The {\it weak $O(3)$ version of the Birkhoff theorem} reads: under the assumption of the spherical symmetry in the sense of invariance under the full rotation $O(3)$ group, when spatial reflections are included along with proper rotations, the Schwarz\-schild (Kottler, in general) spacetime without torsion is a unique solution of the vacuum field equations. In addition to the above cases, the weak theorem holds for the following families in the class of quadratic models (\ref{LRT}): 
\begin{description}
\item[{\it (WB1)}] No curvature square terms, $b_I = 0$ (thus $\Lambda_I = 0$), provided $\mu_2\mu_3 \neq 0$.

\item[{\it (WB2)}] $\Lambda_1 = \Lambda_6 = 0$, $\Lambda_2 \neq 0, \Lambda_3 \neq 0, \Lambda_4 \neq 0, \Lambda_5 \neq 0$, provided the scalar curvature $R(\Gamma)$ is constant, and $\mu^{\rm eff}_2\mu^{\rm eff}_3 \neq 0$.

\item[{\it (WB3)}] $\Lambda_1 = \Lambda_3 = \Lambda_6 = 0$, $\Lambda_2 \neq 0, \Lambda_4 \neq 0, \Lambda_5 \neq 0$, provided $\mu_2\mu_3 \neq 0$. 

\item[{\it (WB4)}] $a_1 = a_2 = 0$, but $a_0 \neq 0$ and $\Lambda_1 = \Lambda_6 = 0$, $\Lambda_2 \neq 0, \Lambda_3 \neq 0, \Lambda_4 \neq 0, \Lambda_5 \neq 0$, provided $a^{\rm eff}_0 \neq 0$.

\item[{\it (WB5)}] $a_1 = a_2 = 0$, but $a_0 \neq 0$ and $\Lambda_6 = 0, \Lambda_1 \neq 0$, $\Lambda_2 \neq 0, \Lambda_3 \neq 0, \Lambda_4 \neq 0, \Lambda_5 \neq 0$, provided the $\xi$ constant (\ref{xill}) satisfies the condition (\ref{exc}).

\item[{\it (WB6)}] $a_1 = a_2 = 0$, but $a_0 \neq 0$ and $\Lambda_1 = \Lambda_6 \neq 0$, $\Lambda_2 \neq 0, \Lambda_3 \neq 0, \Lambda_4 \neq 0, \Lambda_5 \neq 0$, provided the $\xi$ constant (\ref{xill}) satisfies the condition (\ref{exc}).

\item[{\it (WB7)}] $a_1 = a_2 = 0$, but $a_0 \neq 0$ and arbitrary $\Lambda_I$, under the condition of asymptotic flatness. 

\item[{\it (WB8)}] $\Lambda_1 = \Lambda_2 = \Lambda_4 = \Lambda_5 = \Lambda_6 = 0, \Lambda_3 \neq 0$, provided $\mu_1\mu_2\mu_3 \neq 0$ and the curvature scalar vanishing condition. 
\end{description}

\subsection{Correspondence with GR: dynamical torsion beyond Einstein}

The class of Yang-Mills type quadratic Lagrangians (\ref{LRT}) encompasses many interesting and physically viable models. Since Einstein's general relativity theory (GR) is convincingly supported by experiments in terrestrial laboratories and astrophysical observations both in the Solar system and on the extra-galactic scales, it is important to investigate the relation between the general Poincar\'e gravity and GR.

\subsubsection{Einstein-Cartan theory}

The most well known is the so-called Einstein-Cartan theory which is the closest extension of GR. The corresponding Lagrangian is obtained from (\ref{LRT}) by dropping all quadratic terms, for the coupling constants $a_0 = 1$, $a_1 = a_2 = a_3 = 0$, and $b_I = 0$:
\begin{equation}
V_{\rm EC} = -\,{\frac {1}{2\kappa c}}\left(R + 2\lambda_0\right).\label{TEC}
\end{equation}
The gravitational field equations (\ref{PG1})-(\ref{PG2}) then reduce to
\begin{align}
R_i{}^j - {\frac 12}R\delta_i^j - \lambda_0\delta_i^j &= \kappa \Sigma_i{}^j,\label{EC1}\\
T_{ij}{}^k + 2T_{[i}\delta_{j]}^k &= \kappa c\tau_{ij}{}^k.\label{EC2}  
\end{align}
The Einstein-Cartan theory represents a certain degenerate case of the Poincar\'e gauge gravity in the sense that the second field equation (\ref{EC2}) describes an algebraic coupling between the spin of matter and the torsion. This means that the torsion is a non-dynamical field which vanishes outside the matter sources, and thereby the first equation (\ref{EC1}) reduces to Einstein's field equation of GR. 

Resolving (\ref{EC2}), one can express the torsion in terms of the matter spin, and plugging it in (\ref{EC1}), it is possible to recast the latter into an effective Einstein field equation
\begin{equation}
\widetilde{R}_{ij} - {\frac 12}\widetilde{R}g_{ij} - \lambda_0g_{ij}
= \kappa {\stackrel {\rm eff} \Sigma}{}_{ij},\label{EC1eff}
\end{equation}
where the original canonical energy-momentum tensor is replaced by the effective energy-momentum tensor that includes additional contributions of the spin. For the particular case of the spinning fluid (\ref{hypS})-(\ref{hypD}) one finds
\begin{equation}
{\stackrel {\rm eff} \Sigma}{}_{ij} = -\,p^{\rm eff}\Bigl(g_{ij} - {\frac {1}{c^2}}u_iu_j\Bigr)
+ {\frac {\varepsilon^{\rm eff}}{c^2}}\,u_iu_j + \Bigl(g^{kl} + {\frac 1{c^2}}u^k u^l\Bigr)
\widetilde{\nabla}_k\left(u_{(i}{\mathcal S}_{j)l}\right),\label{Seff}
\end{equation}
where the effective pressure and energy density depend on spin:
\begin{align}
p^{\rm eff} &= p -  {\frac {\zeta\kappa c^2}{8}}\,{\mathcal S}_{ij}{\mathcal S}^{ij},\label{QPeff}\\
\varepsilon^{\rm eff} &= \varepsilon -  {\frac {\zeta\kappa c^2}{8}}\,{\mathcal S}_{ij}
{\mathcal S}^{ij}.\label{QEpeff}
\end{align}
Here the numeric constant $\zeta = 1$. 

A qualitatively equivalent model (which can be called a generalized Einstein-Cartan theory ``EC$+$'') is obtained as a natural extension of the Lagrangian (\ref{TEC}) when we include all possible torsion quadratic terms:
\begin{equation}\label{TECG}
V_{\rm EC+} = -\,{\frac {1}{2\kappa c}}\Bigl(R + 2\lambda_0
+ {\frac 12}\sum_{I=1}^3 a_I\,{}^{(I)}\!T^{kl}{}_i\,T_{kl}{}^i\Bigr).
\end{equation}
The resulting field equations then read
\begin{align}
R_i{}^j - {\frac 12}R\delta_i^j - \lambda_0\delta_i^j + T_{in}{}^kh^{jn}{}_k - {\frac 14}\delta_i^j\,T_{mn}{}^kh^{mn}{}_k \nonumber\\
- \,(\nabla_l - T_l)h^{jl}{}_i - {\frac 12}T_{mn}{}^jh^{mn}{}_i &= \kappa \Sigma_i{}^j,\label{ECP1}\\
T_{ij}{}^k + 2T_{[i}\delta_{j]}^k - 2h^k{}_{[ij]} &= \kappa c\tau_{ij}{}^k,\label{ECP2}  
\end{align}
where, recalling (\ref{HT}),
\begin{equation}
h^{ij}{}_k = a_1{}^{(1)}\!T^{ij}{}_k + a_2{}^{(2)}\!T^{ij}{}_k + a_3{}^{(3)}\!T^{ij}{}_k\,.\label{HTEC}
\end{equation}
Since the second field equation (\ref{ECP2}) still describes an algebraic coupling of the matter spin and the torsion, one can resolve the latter and recast (\ref{ECP1}) into an effective Einstein field equation (\ref{EC1eff}). In this model, however, the effective energy-momentum then picks up a dependence on the coupling constants $a_I$. For the case of the spinning fluid (\ref{hypS})-(\ref{hypD}) one again finds (\ref{Seff})-(\ref{QEpeff}), but with a more nontrivial constant
\begin{eqnarray}
\zeta = {\frac {4}{3(1 + a_1)}} - {\frac {1}{3(1 - 2a_3)}}.\label{zetaT}
\end{eqnarray}
It is worthwhile to note that $a_2$ does not contribute in view of the Frenkel condition ${\mathcal S}_{ij}u^j = 0$ imposed on the spin density. When the torsion-square terms are absent, $a_1 = a_2 = a_3 = 0$, we recover the value $\zeta = 1$ of the Einstein-Cartan theory. It is interesting to note that there exists a large class of models with the torsion quadratic Lagrangians (\ref{TECG}) which yield $\zeta = 0$.

Qualitatively, the EC and EC$+$ models are very much alike, because they both can be recast into the form of the effective Einstein theory (\ref{EC1eff}) with the energy-momentum tensor modified by the spin contributions. The magnitude of the terms quadratic in spin in ${\stackrel {\rm eff}\Sigma}{}_{ij}$ becomes comparable with the original canonical energy-momentum tensor $\Sigma_{ij}$ at densities $\rho\ge \rho_{\rm cr} = {\frac {m^2c^4}{\hbar^2 G}}$ of the spinning matter built of particles with the mass $m$ \cite{Hehl:1976kj}. For the mass of a nucleon, the critical density $\rho_{\rm cr} \approx 10^{57}\,$ kg/m$^3$ is still much smaller than the Planck density $\rho_{\rm Pl}\sim 10^{97}\,$ kg/m$^3$ at which the quantum-gravitational effects are expected to start dominating. Consequently, the torsion can be essential already at the level of the classical theory of the gravitational interactions. In particular, this may avert the singularity in the early universe, predicting a finite minimum for the cosmological scale factor reached at the critical matter density \cite{Trautman:1973}.

A thorough analysis of the observational cosmology in the Einstein-Cartan theory with the Weyssenhoff spinning fluid can be found in \cite{Palle:1996,Palle:1999,Brechet:2007,Brechet:2008,Palle:2014}. Other physical consequences of the Einstein-Cartan theory are discussed in great detail in \cite{Hehl:1976kj,Shapiro,Ponomarev}.

\subsubsection{Einstein's GR as a special case of Poincar\'e gravity theory}

In the Einstein-Cartan theory above, we assumed the minimal coupling of the matter to the Poincar\'e gauge fields, which is a natural assumption in the framework of the gauge-theoretic approach.

Quite remarkably, however, one can also view Einstein's general relativity theory as a special case in the framework of the Poincar\'e gauge gravity theory under the assumption of a suitable nonminimal coupling of matter to the Riemann-Cartan geometry of spacetime.

In order to demonstrate this, we start with the extended Einstein-Cartan Lagrangian (\ref{TECG}) and fix the torsion coupling constants as
\begin{equation}\label{V0}
a_1 = -\,1,\qquad a_2 = 2,\qquad a_3 = {\frac 12}.
\end{equation}
As a result, (\ref{HTEC}) reduces to 
\begin{equation}
h^{ij}{}_k = -\,{}^{(1)}\!T^{ij}{}_k + 2{}^{(2)}\!T^{ij}{}_k + {\frac 12}{}^{(3)}\!T^{ij}{}_k
= -\,K_k{}^{ij} - 2T^{[i}\delta^{j]}_k\,,
\end{equation}
where we used (\ref{tT2})-(\ref{tT1}) and (\ref{dist}). A direct computation then yields a considerable simplification of the left-hand sides of the field equations (\ref{ECP1}) and (\ref{ECP2}):
\begin{align}
\widetilde{R}_i{}^j - {\frac 12}\widetilde{R}\delta_i^j &= \kappa \Sigma_i{}^j,\label{EGR1}\\
0 &= \kappa c\tau_{ij}{}^k,\label{EGR2}  
\end{align}
The last equation would be an obviously contradictory relation for the case of the minimal coupling, allowing only for the spinless matter. Nevertheless, this equation becomes meaningful under the assumption of a special type of nonminimal coupling, when the Lagrangian $L_{\rm mat} = L_{\rm mat}(\psi^A, D_i\psi^A, e_i^\alpha, T_{ij}{}^\alpha)$ of the matter fields $\psi^A$ depends on the torsion tensor $T_{ij}{}^\alpha$ that, however, may enter the Lagrangian only in a combination
\begin{equation}\label{Dpsit}
D_i\psi^A - {\frac 12}K_i{}^{\alpha\beta}(\rho_{\alpha\beta})^A{}_B\,\psi^B.
\end{equation}
Then one can demonstrate \cite{Obukhov:2020} that the source on the right-hand side of (\ref{EGR1}) has the form 
\begin{equation}
\Sigma_i{}^j = \overset{\rm m}{\Sigma}{}_i{}^j + {\frac c2}\,\widetilde{\nabla}_k
\Bigl(\overset{\rm m}{\tau}{}^{jk}{}_i + \overset{\rm m}{\tau}{}^j{}_i{}^k
+ \overset{\rm m}{\tau}{}_i{}^{kj}\Bigr),\label{TTc}
\end{equation}
where $\overset{\rm m}{\Sigma}{}_i{}^j$ is the canonical energy-momentum tensor, and  
\begin{equation}
c\overset{\rm m}{\tau}{}_{\alpha\beta}{}^k = (\rho_{\alpha\beta})^A{}_B\,\psi^B{\frac {\partial L_{\rm mat}}
{\partial D_k\psi^A}},\qquad \overset{\rm m}{\tau}{}_{ij}{}^k = 
\overset{\rm m}{\tau}{}_{\alpha\beta}{}^ke^\alpha_ie^\beta_j,\label{Smin}
\end{equation}
is the canonical spin tensor. We immediately recognize in (\ref{TTc}) the well known metrical energy-momentum tensor symmetrized by means of the Belinfante-Rosenfeld procedure.

\subsubsection{Von der Heyde model}

As we saw above, the Einstein-Cartan theory is the closest generalization of Einstein's GR which takes into account the spin of matter as a source of the gravitational field. This changes the geometrical structure of the spacetime manifold, but the torsion remains a non-dynamical field which disappears in the absence of spin. This motivates the study of more general Poincar\'e gravity models in which the torsion becomes a dynamical field. Here we briefly consider two such models. 

The von der Heyde (VdH) model \cite{vdHeyde:1976b,Hehl:EinsteinVolume} attracted considerable attention in the literature. It is described by the Lagrangian that does not contain a linear in the curvature Hilbert-Einstein term, and is purely quadratic in the Poincar\'e gauge field strengths:
\begin{equation}\label{VDH}
V_{\rm VdH} = -\,{\frac {1}{2\kappa c}}\Bigl(- \,{\frac 12}T_{ij}{}^k T^{ij}{}_k + T_iT^i 
+ {\frac {\ell_\rho^2}2}\,R_{ij}{}^{kl}R^{ij}{}_{kl}\Bigr).
\end{equation}
One thus recovers a special case of the general Lagrangian (\ref{LRT}) with $a_0 = 0$, $b_I = 1$, $I = 1,\dots, 6$, and 
\begin{equation}\label{aV}
a_1 = -\,1,\qquad a_2 = 2,\qquad a_3 = -\,1.
\end{equation}
A peculiar feature of the VdH model is that it demonstrates a remarkable compatibility with GR despite the absence of the Hilbert-Einstein term in the Lagrangian. Technically, this is explained by an almost the same set of the torsion coupling constants, cf. (\ref{aV}) and (\ref{V0}).

Unlike the Einstein-Cartan theory, the VdH model predicts nontrivial dynamical torsion effects. To demonstrate this, let us specialize to the spherically symmetric ansatz (\ref{Sds}), (\ref{Scof})-(\ref{Sgam2}). Inspecting the Poincar\'e gauge field equations (\ref{PG1})-(\ref{PG2}) for the Lagrangian (\ref{VDH}), we then find an exact solution for the metric variables
\begin{equation}
A^2 = 1 - {\frac {2m}{r}} + {\frac {r^2}{4\ell_\rho^2}},\qquad
B = {\frac 1A},\qquad C = r,\label{ABCV}
\end{equation}
whereas the anholonomic torsion components $T_{\mu\nu}{}^\alpha = e_\mu^ie_\nu^je^\alpha_kT_{ij}{}^k$ read
\begin{equation}\label{vdhS}
T_{\hat{1}\hat{0}}{}^{\hat{0}} = T_{\hat{1}\hat{0}}{}^{\hat{1}} = T_{\hat{0}\hat{2}}{}^{\hat{2}} =
T_{\hat{2}\hat{1}}{}^{\hat{2}} = T_{\hat{0}\hat{3}}{}^{\hat{3}} = T_{\hat{3}\hat{1}}{}^{\hat{3}} =   
{\frac {m}{A\,r^2}}.
\end{equation}
Here $m = GM/c^2$ is an integration constant, with $M$ interpreted as a total mass of the field configuration. From the point of view of the Riemannian geometry, the line element (\ref{Sds}), (\ref{ABCV}) describes the Schwarzschild-de Sitter (or Kottler) GR solution, where the dynamical torsion induces a ``fake'' cosmological term.

One can extend this result to the axially symmetric case, and demonstrate that the Kerr-de Sitter metric of a massive and rotating field configuration with dynamical torsion is an exact solution in the VdH model \cite{Hehl:EinsteinVolume,Heinicke:2015iva}. Moreover, a systematic analysis \cite{Obukhov:2019} reveals the existence of such solutions in a more general class of Poincar\'e gauge models with the Yang-Mills type Lagrangian (\ref{LRT}). 

\subsubsection{Cembranos-Valcarcel model}

While in VdH model a ``fake'' cosmological term arises from the dynamical torsion, the latter can manifest even more nontrivial effects in the Cembranos-Valcarcel model \cite{Cembranos1,Cembranos2}. The corresponding Lagrangian is a special case of of (\ref{LRT}), where $a_0 = 1$, and the torsion coupling constants are fixed by (\ref{V0}), whereas the curvature coupling sector reads
\begin{equation}\label{CVL}
b_3 = - \,b_2,\quad b_5 = - \,{\frac {b_2}{3}},\qquad b_1 = b_4 = b_6 = 0.
\end{equation}
Specializing again to the spherically symmetric ansatz (\ref{Sds}), (\ref{Scof})-(\ref{Sgam2}), one then finds an exact solution for the metric variables
\begin{equation}
A^2 = 1 - {\frac {2m}{r}} - {\frac {\lambda_0r^2}{3}} + {\frac {Q^2}{r^2}},\qquad
B = {\frac 1A},\qquad C = r,\label{ABCC}
\end{equation}
where an arbitrary integration constant $\sigma_0$ enters
\begin{equation}
Q^2 = {\frac {2b_2\ell_\rho^2\sigma_0^2}3},\label{QCV}
\end{equation}
and determines the structure of the dynamical torsion. The latter (as before, in anholonomic components $T_{\mu\nu}{}^\alpha = e_\mu^ie_\nu^je^\alpha_kT_{ij}{}^k$) reads explicitly:
\begin{align}
T_{\hat{1}\hat{0}}{}^{\hat{0}} = T_{\hat{1}\hat{0}}{}^{\hat{1}} &= {\frac {dA}{dr}},\label{vcS1}\\
T_{\hat{2}\hat{0}}{}^{\hat{2}} = T_{\hat{1}\hat{2}}{}^{\hat{2}} = T_{\hat{3}\hat{0}}{}^{\hat{3}}
= T_{\hat{1}\hat{3}}{}^{\hat{3}} &= {\frac {A}{2r}},\label{vcS2}\\
T_{\hat{0}\hat{3}}{}^{\hat{2}} = T_{\hat{3}\hat{1}}{}^{\hat{2}} = T_{\hat{2}\hat{0}}{}^{\hat{3}}
= T_{\hat{1}\hat{2}}{}^{\hat{3}} &= {\frac {\sigma_0}{A\,r}}.\label{vcS3}
\end{align}
As we now see, from the point of view of the Riemannian geometry, the line element (\ref{Sds}), (\ref{ABCC}) describes the Reissner-Nordstr\"om-de Sitter GR solution, where the dynamical torsion induces a ``fake'' electric charge (\ref{QCV}), and the torsion plays a role of a fictitious electromagnetic field.

An extension of the Cembranos-Valcarcel results for a more general models with both the parity-even and parity-odd sectors included was discussed in \cite{birk}. It is worthwhile to notice that the generalized Birkhoff theorem is not valid for the von der Heyde and the Cembranos-Valcarcel models, and precisely this fact underlies the existence of the spherically symmetric solutions with nontrivial torsion. 

\subsection{Gravitational waves}

In conclusion, it is instructive to discuss a possible physical manifestation of the rich graviton spectrum (\ref{LL}) and (\ref{mumu}) of Poincar\'e gauge gravity theory in the form of the gravitational waves. 

The study of gravitational waves is of fundamental importance in physics, that became an even more significant issue after the purely theoretical research in this area was finally supported by the first experimental evidence \cite{Abbott1,CNN}. The plane-fronted gravitational waves represent an important class of exact solutions \cite{vdz,griff,exact} which generalize the basic properties of electromagnetic waves in flat spacetime to the case of curved spacetime geometry.

Let us discuss the gravitational wave solutions in the PG model with the general quadratic Lagrangian (\ref{LRT}) for the case without the cosmological constant $\lambda_0 = 0$. We start with the flat Minkowski geometry described by the coframe and connection $\widehat{e}{}_i^\alpha = \delta^\alpha_i$, $\widehat{\Gamma}{}_{i\beta}{}^\alpha = 0$, where $x^i = (x^0, x^1, x^2, x^3)$ are Cartesian coordinates. Differentiating the phase variable $\sigma = x^0 - x^1$, we introduce the wave covector $k = \partial_i\sigma = (1, -1, 0, 0)$. With $k_\alpha = \widehat{e}{\,}^i_\alpha k_i$, the gravitational wave ansatz is then introduced as a Kerr-Schild deformation of the flat background:
\begin{align}
e_i^\alpha &= \widehat{e}{}_i^\alpha + {\frac 12}U\,k^\alpha k_i,\label{cofW}\\ 
\Gamma_{i\beta}{}^\alpha &= \widehat{\Gamma}{}_{i\beta}{}^\alpha 
+ (k_\beta W^\alpha - k^\alpha W_\beta)\,k_i.\label{gamW}
\end{align}
The resulting line element (with $\rho = x^0 + x^1$)
\begin{equation}
ds^2 = g_{\alpha\beta}e_i^\alpha e_j^\beta dx^idx^j = d\sigma d\rho + Ud\sigma^2 - \delta_{AB}dx^Adx^B,\label{ds_2}
\end{equation}
represents the plane-fronted wave in the form of Brinkmann \cite{Brink1,Brink3}. By construction, $k_\alpha = (1, -1, 0, 0)$, so that $k^\alpha = (1, 1, 0, 0)$. Therefore, this is a null vector field, $k_\alpha k^\alpha = 0$. 

The gravitational wave configuration (\ref{cofW}) and (\ref{gamW}) is described by the two unknown variables $U$ and $W^a$ which determine wave's profile, they are functions $U = U(\sigma, x^A)$ and $W^\alpha = W^\alpha(\sigma, x^A)$ of the phase $\sigma$ and the transversal coordinates $x^A = (x^2,x^3)$. From now on, the indices from the beginning of the Latin alphabet $a,b,c,\dots = 0,1$, whereas the capital Latin indices run $A,B,C... = 2,3$. In addition, we assume the orthogonality $k_\alpha W^\alpha = 0$, which is guaranteed if we choose 
\begin{equation}\label{Wa0}
W^\alpha = \begin{cases}W^a = 0,\qquad\qquad\qquad a = 0,1, \\
W^A = W^A(\sigma, x^B),\qquad A = 2,3.\end{cases}
\end{equation}

Obviously, $\partial_ik^\alpha = 0$, and $D_ik^\alpha = 0$ for the wave covector with constant components, and we straightforwardly find the torsion and the curvature:
\begin{eqnarray}
T_{kl}{}^i = -\,2k^ik_{[k}\Theta_{l]},\qquad R_{kl}{}^{ij} = -\,4k_{[k}k^{[i}\Omega_{l]}{}^{j]}\,.\label{curW}
\end{eqnarray}
Here we constructed the two objects from the derivatives of $U = U(\sigma, x^A)$ and $W^\alpha = W^\alpha(\sigma, x^A)$ with respect to the transversal coordinates $x^A = (x^2, x^3)$:
\begin{align}\label{TMW}
\Theta_i &= \Bigl\{\Theta_a = 0,\quad \Theta_A = {\frac 12}\,\partial_AU -\delta_{AB} W^B\Bigr\},\\
\Omega_j{}^i &= \Bigl\{\Omega_b{}^a = 0,\ \Omega_B{}^a = 0, \ \Omega_b{}^A = 0,\
\Omega_B{}^A = \partial_BW^A\Bigr\}. \label{OMW}
\end{align}

The translational and rotational Poincar\'e gauge field strengths (\ref{curW}) have qualitatively the same structure as the electromagnetic field strength $F_{ij}$ of a plane wave, that has the properties $k^jF_{ij} = 0$, $k_{[i}F_{jk]} = 0$, $F_{ij}F^{ij} = 0$. In complete analogy, the Poincar\'e gauge field strengths of a gravitational wave satisfy
\begin{alignat}{3} \label{kTW}
k^jT_{ij}{}^k &= 0,\qquad & k_{[i}T_{jk]}{}^l &= 0,\qquad & T_{ij}{}^kT^{ij}{}_l &= 0,\\
k^jR_{ij}{}^{kl} &= 0,\qquad & k_{[i}R_{jk]}{}^{mn} &= 0,\qquad & R_{ij}{}^{kl}R^{ij}{}_{mn} &= 0.\label{kRW}
\end{alignat}
In addition, however, for the gravitational Poincar\'e gauge field strengths we find
\begin{equation}
k_lT_{ij}{}^l = 0,\qquad k_lR_{ij}{}^{kl} = 0.\label{kTRW2}
\end{equation}

The torsion (\ref{curW}) vanishes when $\Theta_i = 0$, which means that $W^A = {\frac 12}\delta^{AB}\partial_BU$, Then $U$ remains the only nontrivial variable and the solution reduces to the usual plane gravitational wave of the Riemannian GR. By noticing this, it is convenient to express the wave profile vector variable in terms of potentials
\begin{equation}\label{pW}
W^A = {\frac 12}\delta^{AB}\partial_B(U + V) + {\frac 12}\eta^{AB}\partial_B\overline{V},
\end{equation}
where $\eta_{AB} = -\,\eta_{BA}$ is the totally antisymmetric Levi-Civita tensor on the 2-dimensional space of the wave front. This brings us to the physically transparent representation of the plane wave in the  Poincar\'e gauge gravity in terms of the three scalar variables $U = U(\sigma, x^A)$, and $V = V(\sigma, x^A)$, $\overline{V} = \overline{V}(\sigma, x^A)$, where the first one is a Riemannian mode, and the two last ones account for the torsion wave modes. 

The explicit gravitational wave solution is constructed as follows \cite{YNOwave1,YNOwave2}. Substituting the wave ansatz (\ref{cofW}), (\ref{gamW}) and (\ref{pW}) into the gravitational field equations (\ref{PG1}) and (\ref{PG2}), the latter reduce to the system of three linear differential equations
\begin{align}
a_0\,\Delta\,U - \mu_3\,\Delta\,V &= 0,\label{UW}\\
\ell_\rho^2\,\Lambda_1(a_0 + \mu_3)\,\Delta\,V - a_0\mu_3\,V &= 0,\label{VW}\\
\ell_\rho^2\,\Lambda_2\,\Delta\,\overline{V} - \mu_3\,\overline{V} &= 0.\label{UVW}
\end{align}
Here $\Delta = \delta^{AB}\partial_A\partial_B$ is the 2-dimensional Laplacian on the $(x^2,x^3)$ space. Using the solutions of (\ref{VW}) and (\ref{UVW}) for the torsion waves in (\ref{UW}), we can find the Riemannian mode $U$ from the resulting inhomogeneous equation.

Quite remarkably, all the three wave modes are massless when $\mu_ 3 = 0$.

\section{Teleparallel gravity}

Presently, considerable research efforts are focused on the teleparallel theory of the gravitational field. From the gauge-theoretic point of view, the latter is based on the gauging of the group of spacetime translations, and it is worthwhile to mention that the fundamental relation of translation symmetry to gravity was clear already at the beginning of the 1960s to Sakurai, Glashow, Gell-Mann, and Feynman (see the historic account in \cite{Hehl:2020}). The conserved energy-momentum current of matter is associated to translations via the Noether theorem, and it naturally arises as a physical source of the corresponding gauge gravitational field.  

\begin{figure}
\centering
\includegraphics[width=0.55\textwidth]{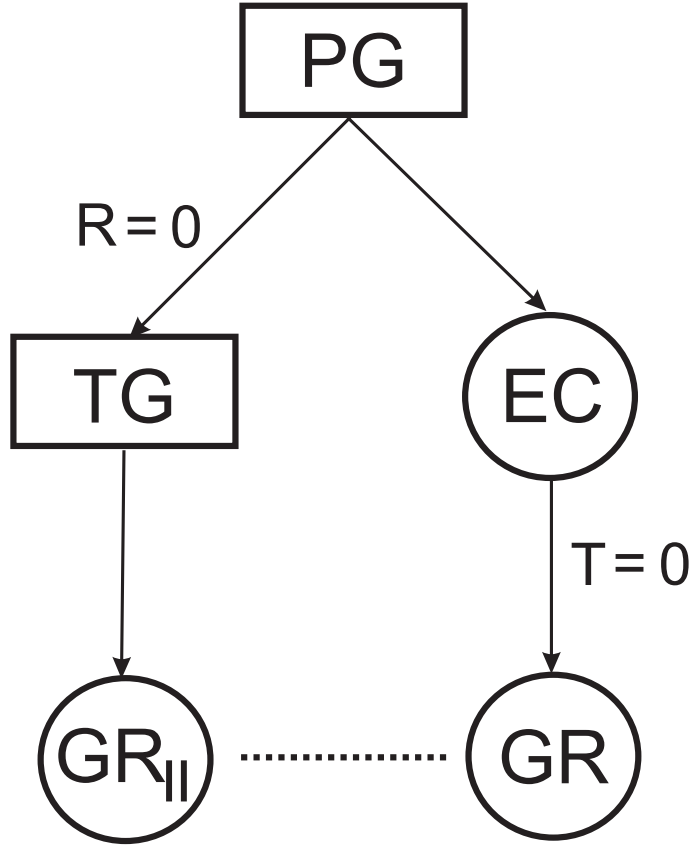}
\caption{Classification of Poincar\'e gauge theories of gravity (see the frontispiece of \cite{Blagojevic:2013}): {\bf PG} = Poincar\'e gauge gravity, {\bf EC} = Einstein-Cartan theory, {\bf GR} = Einstein's general relativity, {\bf TG} = translation gauge theory (teleparallel theory), {\bf GR}$_{||}$ = a specific TG known as teleparallel equivalent of GR. The symbols denote here: rectangle $\Box$ -- general class of theories; circle ${\bigcirc}$ -- viable models.}\label{RCmod}
\end{figure}

The structure of the teleparallel gravity (TG) as a translational gauge theory became essentially established since 1970s, see \cite{Hay1,Cho:1975dh,Nitsch:1979qn,Hay2,Obukhov:2002tm,Pereira:2019,Koivisto:2019,Aldrovandi:2013,Maluf:2012,Maluf:2013,Itin:2001,Hehl:2016glb,Itin:2018dru,Itin:2016nxk,illumi}. A revival of interest to the gauge-theoretic subtleties underlying TG has lead to a recent highly enlightening discussion \cite{Obukhov:2002tm,Fontanini,Delliou,Huguet1,Huguet2}, in particular within the fruitful framework of Tartu conferences. Since the teleparallel gravity theory is considered in full depth in the comprehensive review of Manuel Hohmann in this volume, here we merely highlight the main features of TG as a special case of PG, see Fig.~\ref{RCmod}.

The gauging of the group of translations yields the condition (\ref{Wei}) that introduces the distant parallelism geometry of Weitzenb\"ock on the spacetime manifold. 

As a result, the general Yang-Mills type Lagrangian (\ref{LRT}) reduces to (we confine attention to the parity-even case, and assume zero cosmological constant, for now) \cite{Itin:2016nxk,Pellegrini,Kaempfer}
\begin{equation}\label{TG}
V_{\rm TG} = -\,{\frac {1}{4\kappa c}}\left(a_1\,{}^{(1)}\!T^{kl}{}_i\,T_{kl}{}^i
+ a_2\,{}^{(2)}\!T^{kl}{}_i\,T_{kl}{}^i + a_3\,{}^{(3)}\!T^{kl}{}_i\,T_{kl}{}^i\right). 
\end{equation}
The gravitational field equations can be derived from the action principle either by implementing the teleparallel condition (\ref{Wei}) by means of the Lagrange multiplier, or by making use of the gauge $\Gamma_{i\alpha}{}^\beta = 0$ which means that the connection (\ref{GG2}) takes the Weitzenb\"ock form
\begin{equation}
\Gamma_{ki}{}^j = e^j_\alpha\partial_ke^\alpha_i.\label{Gwei}
\end{equation}
Then the torsion reduces to the anholonomity object (\ref{anhol}).

The dynamical contents of a general TG model (\ref{TG}) strongly depends on the values of the coupling constants $a_1, a_2, a_3$. In particular, in generic case, black hole solutions are absent in this theory \cite{Hay2,Obukhov:2002tm}, and there is no consistency with GR. 

However, for a very special case when the coupling constants $a_1, a_2, a_3$ take the values (\ref{V0}), the dynamics of the gravitational field is fully consistent with Einstein's theory. Then the Lagrangian is simplified to
\begin{equation}\label{TGII}
V_{{\rm GR}_{||}} = -\,{\frac 1{2\kappa c}}\Bigl(-\,{\frac 14}\,T_{kl}{}^i\,T^{kl}{}_i
+ T_i\,T^i + {\frac 12}\,T_{kl}{}^i\,T_i{}^{kl}\Bigr),
\end{equation}
and this model is called a {\it teleparallel equivalent of GR}. 

\section{Conclusion and outlook}

In this review we presented, at an elementary level using the standard tensor language, the formulation of the theory of gravitational interaction as a gauge theory of the Poincar\'e symmetry group. This approach is developed along the lines of a heuristic scheme in which a new physical interaction is derived in the Lagrange-Noether formalism from a conserved current corresponding to the rigid symmetry group by extending the latter to a local symmetry. Leaving aside the derivation of the relevant conservation laws, which was thoroughly discussed earlier in \cite{Obukhov:2006,Obukhov:2018,Hehl:2020}, we have formulated here the general dynamical scheme of Poincar\'e gauge gravity for the class of the Yang-Mills type models (\ref{LRT}) and considered a selected number of particular physically interesting models.

A more mathematically elaborated formulation of the gauge gravity approach in terms of the modern differential geometry language of the affine frame bundle can be found in \cite{PRs,Mielke,Ponomarev,Hehl:deBroglie}. We did not intend to give a detailed review of the physical contents of the Poincar\'e gauge gravity theory. This subject was intensively studied in the past and the relevant results are available in the classic reviews \cite{Hehl:1976kj,Hehl:EinsteinVolume,Shapiro}. At present we are again observing a considerable growth of interest to the gauge gravitational issues. The search and analysis of exact solutions of the gravitational field equations is at the center of the current research, which is essential for improvement of understanding of the nature of the gravitational interaction \cite{Obukhov:1989,Cembranos1,Cembranos2,Heinicke:2015iva,Obukhov:2019,Spindel:2021}.

It is worthwhile to mention that the recent advances in the modern cosmological science have seriously warmed up the interest in the thorough revision of universe's evolution in the broad framework of the modified gravity theories and, in particular, in Poincar\'e gauge gravity. The early predictions \cite{Trautman:1973,Mink1,Mink2} of a possible avertion of singularity in the early universe, and more recent proposals of possible modifications of the late stage of cosmological evolution \cite{Magu,Magu2,Pop,Pop2,Pop3,Dirk}, are currently revisited and extended with an aim to better understand the role of the torsion in the early universe and to resolve the problem of the dark energy \cite{Zhang:2019,Kranas:2018,Barrow:2019,Ivanov:ast}, furthermore, the inclusion of parity-odd sector was critically evaluated in \cite{Baekler1,Baekler2,Chen,Ho2,Ho3,Ho4}.

The last but not least remarks are in order about the direct experimental tests and estimates of the torsion effects to probe possible deviations of the spacetime structure beyond the Riemannian geometry, in accordance with Einstein's \cite{Einstein:1921GE} statement that ``...the question whether this continuum has a Euclidean, Riemannian, or any other structure is a question of physics proper which must be answered by experience, and not a question of a convention to be chosen on grounds of mere expediency.'' The consistent analysis \cite{vdHeyde:1976b,Yasskin,Hehl:2013,Obukhov:2015eqa} of the propagation equations, derived from the conservation laws of PG theory in the framework of the multipole expansion approach, demonstrates that the torsion couples only to the intrinsic spin and never to the orbital angular momentum of test particles. The predicted spin-torsion effects are expected to be quite small, and no spacetime torsion effects were directly observed so far. From the analysis of the data available from precision experiments with spinning particles in high energy physics and astrophysical observations, performed in the numerous theoretical studies \cite{Hehl71,Adamowicz,AudLam,Rumpf,Aud,Lam:1997,Ni2010,Alan1,Alan2,Lehnert,Ivanov2,Ivanov4,ostor,Tru}, one typically finds a rather strong bound $|T| \lesssim 10^{-15}\,$m$^{-1}$ for the magnitude of the spacetime torsion. 

\subsection*{Acknowledgments}
I am grateful to Friedrich Hehl for the careful reading of the manuscript and helpful comments.

\end{document}